\RequirePackage{etex} 
\RequirePackage{fix-cm}
\documentclass[smallextended]{svjour3}
\smartqed 

\usepackage{graphicx}%
\usepackage{multirow}%
\usepackage[title]{appendix}%
\usepackage{xcolor}%
\usepackage{textcomp}%
\usepackage{manyfoot}%
\usepackage{booktabs}%
\usepackage{algorithm}%
\usepackage{algorithmicx}%
\usepackage{algpseudocode}%
\usepackage{listings}%
\usepackage{lmodern}%

\usepackage{amsmath,amssymb,amsfonts,nicefrac}%
\usepackage{mathrsfs}%

\newcommand{\bone}{\mathbf{1}}
\newcommand{\R}{\mathbb{R}}

\usepackage{microtype}
\usepackage{booktabs}
\usepackage[dvipsnames]{xcolor}
\usepackage[colorlinks]{hyperref}
\hypersetup{
	linkcolor=BrickRed
	,citecolor=Green
	,filecolor=Mulberry
	,urlcolor=NavyBlue
	,menucolor=BrickRed
	,runcolor=Mulberry
	,linkbordercolor=BrickRed
	,citebordercolor=Green
	,filebordercolor=Mulberry
	,urlbordercolor=NavyBlue
	,menubordercolor=BrickRed
	,runbordercolor=Mulberry
}

\usepackage{tikz}
\usepackage{tikz-network}
\usetikzlibrary{matrix,positioning,external} 
\usepackage{pgfplots}
\pgfplotsset{compat=1.18}
\usepackage[caption=false]{subfig}
\usepackage{siunitx}

\newcommand{\NysTrace}{\textsc{XNysTrace}\,}
\newcommand{\NysTraceExp}{\textsc{XNysTrace-exp}\,}

\journalname{Journal of Scientific Computing}

\begin{document}

\title{Walk based Laplacians for Modeling Diffusion on Complex Networks\thanks{This work has been partially supported by:  Spoke 1 ``FutureHPC \& BigData''  of the  Italian Research Center on High-Performance Computing, Big Data and Quantum Computing (ICSC)  funded by MUR Missione 4 Componente 2 Investimento 1.4: Potenziamento strutture di ricerca e creazione di ``campioni nazionali di R\&S (M4C2-19 )'' - Next Generation EU (NGEU). The visit of F.D. at the University of Strathclyde where some of this research was carried out was funded by  ``Istituto Nazionale di Alta Matematica (INdAM) - Gruppo Nazionale per il Calcolo Scientifico (GNCS)'' project ``PASTRAMI - sPline And Solver innovaTions foR Adaptive isogeoMetric analysIs". Project code CUP E53C24001950001.
F.D. acknowledges the MUR Excellence Department Project awarded to the Department of Mathematics, University of Pisa, CUP I57G22000700001.
}
}

\titlerunning{Walk based Laplacians}        %

\author{Francesca Arrigo         \and
        Fabio Durastante %
}

\authorrunning{F. Arrigo, F. Durastante} %

\institute{F. Arrigo \at
              Department of Mathematics and Statistics \\
              University of Strathclyde \\
              26 Richmond Street, Glasgow, G1 1XH, Glasgow City, Scotland \\
              \email{francesca.arrigo@strath.ac.uk}  \\         %
              ORCID: 0000-0001-5473-7284 
           \and
           F. Durastante \at
           Department of Mathematics \\
           University of Pisa \\
           Largo Bruno Pontecorvo 5, Pisa, 56127, PI, Italy \\
           \email{fabio.durastante@unipi.it} \\
           ORCID: 0000-0002-1412-8289
}

\date{Received: \today}

\maketitle

\begin{abstract}
We develop a novel framework for modeling diffusion on complex networks by constructing Laplacian-like operators based on walks around a graph. Our approach introduces a parametric family of walk-based Laplacians that naturally incorporate memory effects by excluding or downweighting backtracking trajectories, where walkers immediately revisit nodes. The framework includes: (i) walk-based Laplacians that count all traversals in the network; (ii) nonbacktracking variants that eliminate immediate reversals; and (iii) backtrack-downweighted variants that provide a continuous interpolation between these two regimes. We establish that these operators extend the definition of the standard Laplacian and also preserve some of its properties. We present efficient algorithms using Krylov subspace methods for computing them, ensuring applicability of our proposed framework to large networks. Extensive numerical experiments on real-world networks validate the modeling flexibility of our approach and demonstrate the computational efficiency of the proposed algorithms, including GPU acceleration.
\end{abstract}

\maketitle

\section{Introduction}

Understanding diffusion and information spread in complex networks is fundamental in numerous applications, ranging from epidemiology and social networks to transportation systems and biological processes; see~\cite{DIFFSURVEY} and references therein. The graph Laplacian operator $L$ is a widely used tool for modeling such phenomena, with its classical definition only incorporating nearest-neighbor interactions. 
Over the past decade, several new extensions have been developed in order to capture longer-range interactions and incorporate memory effects in the exploration of networks~\cite{DIFFSURVEY}. These extensions can be broadly classified into two families; top-down approaches construct non-local diffusion operators through fractional powers of the standard graph Laplacian (e.g., $L^\gamma$ with $\gamma \in (0,1)$)~\cite{MR4130854,Riascos2,Riascos1}, enabling super-diffusive and sub-diffusive behaviors; and bottom-up approaches which, on the other hand, explicitly construct transition matrices based on the enumeration of specific path structures~\cite{MR4216832,estrada,MR2900722,MR3834211}, offering greater interpretability and flexibility in modeling network exploration. 
Both these approaches, however, have an extremely high computational cost; indeed, fractional Laplacians require the explicit construction of matrix functions~\cite{MR2396439} whereas path-based approaches demand costly all-pairs shortest-path computations~\cite{Floyd,Bernard,Warshall}. 

The key innovation of this work is the development of a flexible bottom-up framework built on \emph{walks} rather than paths. This immediately yields computational advantages while at the same time allowing fine control over diffusion dynamics through the incorporation of memory effects. Our framework further encompasses a parametric family of Laplacian-like operators that naturally exclude or downweight backtracking trajectories, ensuring that walkers are penalised for immediately revisiting nodes they have just left. This enables us to capture realistic exploratory behavior on networks and introducing memory effects in our model. 

Our approach unifies and extends several existing concepts. Further, it incorporates the use of nonbacktracking walks, that have so far mainly been studied for centrality measures~\cite{MR3801714,MR3842589,MR3769701}. Their systematic application to diffusion operators and the incorporation of graded backtrack-downweighting~\cite{MR4296758} are novel contributions.

The proposed framework retains the same interpretabiliy of previously introduced operators defined in terms of enumerating paths~\cite{MR4216832,estrada,MR2900722,MR3834211}. However, the family of operators we introduce offers several advantages when compared to those; 
(i) the operators naturally incorporate memory through nonbacktracking and backtrack-downweighted formulations, providing a parametric interpolation between standard diffusion to purely nonbacktracking dynamics; 
(ii) the matrix functions appearing in our definitions {do not need to be} built, as their action over a vector can be efficiently computed via Krylov subspace methods without assembling dense matrices. This in turn ensures scalability and applicability of our framework to large networks; and 
(iii) the diffusion equations induced by the new operators preserve properties of the standard Laplacian, thus ensuring theoretical consistency.

The remainder of the paper is organized as follows. Section~\ref{sec:notation_and_bg} presents notation and relevant background on graph theory and the classical diffusion equation. Section~\ref{sec:kwalkdiffusion} develops the main framework, introducing $k$-walk Laplacians and their backtrack-downweighted variants, establishing their mathematical properties and interpretations through their connection with network centrality concepts. Section~\ref{sec:computational_issues} addresses computational challenges and proposes efficient algorithms for the computation of matrix-vector products, trace estimation, and linear system solutions tailored to these new operators. Section~\ref{sec:numerical_experiments} validates the approach through numerical experiments on real-world networks, comparing diffusion behavior, computational scaling, and the effect of backtrack-downweighting. 

\section{Background and Notation}\label{sec:notation_and_bg}

One of the most common ways of representing complex networks is by means of graphs. Formally, an \emph{unweighted graph} is defined as an ordered pair of sets $G = (V, E)$, where $V = \{1,2,\ldots,n\}$ is the set of \emph{nodes} and $E \subseteq V \times V$ is the set of \emph{edges}. 
Henceforth, the edge $(i,j)\in E$ from node $i$ to  node $j$ will be equivalently represented using the notation $i\to j$. 
Whenever $E$ is a set (i.e., it does not contain repeated elements) that is symmetric (i.e., $(i,j)\in E \Longleftrightarrow (j,i)\in E$) and antireflective (i.e., $(i,i)\not\in E$ for all $i\in V$), $G$ will be called \textit{simple}. 
Throughout this work, we will only consider simple unweighted graphs. 

We are interested in linear algebraic representations of graphs. Perhaps the most widely used matrix representation of networks is the adjacency matrix.
\begin{definition}%
\label{def:adjacency}
Let $G = (V,E)$ be an unweighted graph with $n$ nodes. Its \textit{adjacency matrix} $A \in \mathbb{R}^{n \times n}$ is entrywise  defined as:
\[ A_{ij}=\begin{cases}
	1 \ \ \text{if}\,\, (i,j) \in E\\
	0 \ \ \text{otherwise}
\end{cases} \quad \forall\,i,j = 1,2,\ldots, n.
\]
\end{definition}

When the underlying graph $G$ is simple the adjacency matrix is symmetric with zeros on the main diagonal. 

\medskip 

The constructions that we will present in the next sections all build on the idea of walks around a graph. 
\begin{definition}%
\label{def:walks_and_paths}
	A \textit{walk of length $k$} from $i$ to $j$ is a sequence of $k+1$ nodes $i_1,i_2,\ldots,i_{k+1}$ such that $i_1 = i$, $i_{k+1} = j$, and $(i_\ell,i_{\ell+1})\in E$ for all $\ell=1,2,\ldots, k$. If $i_\ell\neq i_\kappa$ for all $\ell\neq\kappa$, then the walk is called a \textit{path}.
\end{definition}
While edges describe immediate connections among nodes, the concept of walk can be used to analyse long range interactions between pairs of nodes. This is showcased by the following classical result. 
\begin{proposition}[{\cite[Ex. 34, Chapter~1]{MR2159259}}]\label{pro:power-of-adjacency}%
Let $A$ be the adjacency matrix of an unweighted graph $G = (V, E)$. Then, the number of walks of length $k\in\mathbb{N}$ from node $i$ to node $j$ is given by $(A^k)_{ij}$ for all $i,j\in V$.
\end{proposition}
An immediately related concept is that of connectivity of a simple graph. 
\begin{definition}%
A simple graph is deemed \textit{connected} if there exists a walk connecting all its nodes and \textit{disconnected} otherwise. Whenever $G$ is disconnected, its maximal connected subgraphs are referred to as its \textit{connected components}.
\end{definition}
In linear algebraic terms,  a connected graph $G$ will correspond to an irreducible adjacency matrix $A$.

The idea of path, paired with Proposition~\ref{pro:power-of-adjacency}, allows one to define a concept of distance between nodes in connected graphs. More specifically, the \textit{geodesic} or \textit{shortest path distance} between nodes $i$ and $j$ in a simple connected graph is defined as the length of the shortest path connecting the two nodes, i.e., 
\[
\operatorname{d}(i,j) = \min\{k\in\mathbb{N}|\,(A^k)_{ij}>0\}.
\]

Without loss of generality, from this point on we will assume that the graphs under consideration are unweighted, simple, and connected. Whenever a graph is disconnected, our theory can be independently applied to the individual connected components. We will also assume the following notation; $I\in\mathbb{R}^{n\times n}$ is the identity matrix, $O\in\mathbb{R}^{n\times n}$ is the matrix of all zeros, $\mathbf{0},\bone\in\mathbb{R}^n$ are the (column) vector of all zeros and all ones, respectively, $\mathbf{e}_i$ is the $i$th (column) vector of the standard basis of $\mathbb{R}^n$, and $\|\cdot\|$ denotes the Euclidean norm. {Further, $\sigma(M)$ and $\rho(M)$ denote the spectrum and spectral radius of a matrix $M \in \mathbb{R}^{n \times n}$, respectively}. 

\subsection{The graph Laplacian and the diffusion equation}\label{sec:laplacian_matrix_and_diffusion_equation} 

The (combinatorial) graph Laplacian is a fundamental matrix in graph theory and spectral graph theory. 
\begin{definition}%
\label{def:undirected_laplacian}
Let $G = (V,E)$ be a simple graph with adjacency matrix $A$.
We define the {Laplacian matrix} as 
\[
L = D - A,
\]
where $A$ is the adjacency matrix of $G$ and $D\in\mathbb{R}^{n\times n}$ is the degree matrix, a diagonal matrix with $D_{ii} = (A\bone)_i$ for all $i\in V$.
\end{definition}

{The graph Laplacian is a particular instance of a more general matrix
construction that will be useful throughout the paper; see Lemma~\ref{lem:being_an_m_matrix} below. We first recall the
following definitions~\cite{MR1298430}.
\begin{definition}
A matrix $M\in\mathbb{R}^{n\times n}$ is called a \textit{$Z$-matrix} if all its
off-diagonal entries are nonpositive, i.e.,
$M_{ij}\leq0$ for all $i\neq j$.
\end{definition}
\begin{definition}
A matrix $M\in\mathbb{R}^{n\times n}$ is called an \textit{M-matrix} if it is a
$Z$-matrix and all its eigenvalues have nonnegative real parts, i.e.,
\[
\operatorname{Re}(\lambda)\geq0
\qquad\text{for all }\lambda\in\sigma(M).
\]
\end{definition}
We note in passing that several equivalent properties can be used to show that a given Z-matrix is an M-matrix; see, e.g., \cite[Theorem 2.5.3]{roger1994topics}.

\begin{lemma}\label{lem:being_an_m_matrix}
Let $C\in\mathbb{R}^{n\times n}$ be a symmetric nonnegative matrix. 
Then, the matrix $M=\operatorname{diag}(C\mathbf{1})-C$ is a symmetric, singular M-matrix and, moreover, $M\mathbf{1}=\mathbf{0}$.
\end{lemma}
\begin{proof}
Symmetry of $M$ holds true by construction, since $C=C^\top$. 
Moreover, by construction, since $C\geq0$, the off-diagonal entries of $M$ satisfy
$M_{ij}=-C_{ij}\leq0$ for $i\neq j$. Thus $M$ is a symmetric $Z$-matrix and $\sigma(M)\subset \R$. 
We now show that $\sigma(M)\subseteq [0,\infty)$, hence that $M$ is an M-matrix, by proving that $M$ is positive semidefinite. To this end, consider any vector 
$\mathbf{x}\in\mathbb{R}^n$. Exploiting the definition of $M$ and symmetry of $C$, it can be shown that  
$2\mathbf{x}^{\top}M\mathbf{x}%
=\sum_{i,j=1}^nC_{ij}(x_i-x_j)^2$,
which is nonnegative. Thus $M$ is positive semidefinite. 
Finally, it is readily verified that  $M\mathbf{1}
=(\operatorname{diag}(C\mathbf{1})-C)\mathbf{1}
=\operatorname{diag}(C\mathbf{1})\mathbf{1}-C\mathbf{1}
=\mathbf{0}$.
\end{proof}}

{Taking $C=A$ in Lemma~\ref{lem:being_an_m_matrix} immediately gives that
the combinatorial graph Laplacian $L=\operatorname{diag}(A\mathbf{1})-A$ is a singular
M-matrix. 
For the purpose of this work, the relevance of $L$ being an M-matrix lies in its relation to the diffusion equation on a graph. Indeed, non-singular M-matrices are inverse positive. As such, they ensure that the discrete diffusion model inherently respects a maximum principle, physically guaranteeing that the evolution smoothly transports quantities from high concentrations to low concentrations, without producing negative values or spurious oscillation. 
For singular M-matrices this corresponds to a closed system where mass (or thermal energy) is strictly conserved; the physical process will smooth out spatial gradients into a uniform, non-zero equilibrium state preserving the initial total mass (or thermal energy). 
Therefore, the evolution under an M-matrix dynamic mimics the behavior of the continuous diffusion operator which satisfies the continuous maximum principle; we refer the interested reader to~\cite{MR292317} for a detailed discussion of the connection between the continuous and discrete maximum principles.}

Given a row vector $\mathbf{p}_0\geq 0$ with $\mathbf{p}_0 \mathbf{1} = 1$, the diffusion equation is
\begin{equation}\label{eq:diffusion_on_G}
\begin{cases}
\displaystyle \frac{\mathrm{d} \mathbf{p}(t)}{\mathrm{d} t} = - \mathbf{p}(t) L, & t \in [0, t^*], \\
\mathbf{p}(0) = \mathbf{p}_0, 
\end{cases}
\end{equation}
which is solved, for all times $t$, by the probability distribution 
\[
\mathbf{p}(t) = \mathbf{p}_0 \exp(- t L).
\]

This version of the diffusion equation is often referred to as \emph{Poissonian edge centric}~\cite{DIFFSURVEY}, since the transition rate to a given node is determined by its degree; a walker tends to leave high-degree nodes faster than low-degree ones. 

Since {the combinatorial Laplacian $L$} is a {singular} M-matrix, {its eigenvalues are all nonnegative.}
Hence, the negative sign in~\eqref{eq:diffusion_on_G} ensures that the substance diffusing on the graph flows from regions of higher concentration to lower concentration, mirroring the behavior of heat diffusion in physical systems. The eigenvalues and eigenvectors of $L$ significantly influence the behavior of the diffusion process, particularly in characterizing the long-term dynamics and equilibrium states of~\eqref{eq:diffusion_on_G}, which is readily shown to be collinear to $\bone$. Indeed, the interesting part of the diffusion phenomena is the transient phase.

\section{Going for a stroll: \texorpdfstring{$k$-walk Laplacians}{k walk Laplacians} and their backtrack-downweighted version}\label{sec:kwalkdiffusion}

The Laplacian matrix $L$ is defined by only considering immediate connections between nodes in the graph, and therefore it captures primarily local phenomena. Recent literature has shown a growing interest in incorporating longer-range interactions within the definition of variants of the Laplacian. 
The approaches to doing this can be broadly categorized into two groups; a top-down approach, exemplified by the use of the fractional Laplacian $L^\gamma$ with $\gamma \in (0,1)$, which generalizes the discrete Laplacian $L$ through the use of a matrix function~\cite{Riascos1,MR4130854}, and a bottom-up approach, where the paths in the graph are utilized in the definition of new measures~\cite{MR2900722,MR3834211,Riascos2}. 
In the following, we will construct a bottom-up approach that considers walks instead of paths. 
This will allow us to define a parametric family of Laplacians that on the one side capture a walker's short-term memory, and on the other allow to perform long jumps on the graph. These jumps will occur between two nodes $i$ and $j$ with a probability depending not only on the length of the walks connecting $i$ and $j$, but also on the number of such walks.

We begin by recalling the construction of the \textit{$k$-path transformed Laplacian} $\mathcal{L}$, which was introduced in~\cite{MR2900722,MR3834211} as an alternative to the fractional powers of $L$. In \cite{MR3834211} the authors propose to build a sequence of Laplacians, each encoding information on paths occurring between nodes of a specific length only, if any. 
If we let $\operatorname{diam}(G) = \max\{\mathrm{d}(i,j)|\,i,j\in V\}$ be the \textit{diameter} of the graph, i.e., the length of the longest shortest path in $G$, then, for each $k= 1,2,\ldots, \operatorname{diam}(G)$ the associated \textit{$k$-path Laplacian}  $\mathcal{L}_k$ of $G$ is defined entrywise as follows;
\begin{equation}\label{eq:kpathlaplacian}
\left(\mathcal{L}_k\right)_{ij} = 
\begin{cases}
\mathcal{N}_k(i), & \text{if } i = j,\\
-1, & \text{if } \mathrm{d}(i,j) = k,\\
0, & \text{otherwise},
\end{cases} 
\end{equation}
where $\mathcal{N}_k(i) = \left\lvert \{ j \in V \mid \mathrm{d}(i,j) = k \} \right\rvert$ is the number of nodes at distance $k$ from node~$i$. For $k>\operatorname{diam}(G)$, $\mathcal{L}_k=O$. 
Clearly, this definition extends Definition~\ref{def:undirected_laplacian} in the sense that these matrices retain the ``idea" of how $L$ is built (``degrees" on the diagonal and~$-1$ in selected off-diagonal entries). 

The \textit{$k$-path transformed Laplacian}, which is predominantly used in practice, is then defined as;
\begin{equation}\label{eq:kpathlaplacian-global}
\mathcal{L} = \sum_{k=1}^{\infty} t_k \mathcal{L}_k= \sum_{k=1}^{\operatorname{diam}(G)} t_k \mathcal{L}_k, \qquad t_k \in \left\lbrace \frac{1}{k^\beta}, \exp(-\beta k) \right\rbrace_{\beta \in \mathbb{R}_+,\,k \in \mathbb{N}}.
\end{equation}
A striking limitation of this approach is that it requires computing and storing the distances between any two nodes in the graph, a computationally expensive and memory-taxing procedure; see Section~\ref{sec:computational_issues} for more details.

\medskip

Because of this limitation, and because of how much simpler and cheaper it is to count walks rather than paths, we seek to introduce the alternative definition  of \textit{$k$-walk Laplacians} and of the associated \textit{$k$-walk transformed Laplacian}; see Definitions~\ref{def:kwalklaplacian} and \ref{def:transformedlaplacian} below.
This will be achieved by exploiting Proposition~\ref{pro:power-of-adjacency}. 
From the modeling viewpoint, for any $k\in\mathbb{N}$ a walker will move from node $i$ to node $j$ on the graph  with a weight that is proportional to the number of walks of length $k$ between the two.

\begin{definition}\label{def:kwalklaplacian}
For any $k\in\mathbb{N}$, the \textit{$k$-walk Laplacian} is defined as 
\begin{equation}\label{eq:kwalklaplacian}
\mathbb{L}_k = \operatorname{diag}(A^k \mathbf{1}) - A^k.
\end{equation}
\end{definition}
We point out that, differently from their path-based equivalents $\mathcal{L}_k$, these matrices may be dense even for sparse graphs. The sparsity (or otherwise) of $\mathbb{L}_k$ will depend on the connectivity structure of the graph; for example, whenever $A$ is primitive\footnote{An adjacency matrix is said to be primitive if it is irreducible and has only one eigenvalue of maximum modulus; see, e.g.,~\cite[Definition 8.5.0, Theorem 8.5.2]{MR2978290}.} there exists $k>0$ such that $A^k>0$.
Still, the computation of $\mathbb{L}_k$ is much simpler and cheaper. In fact, in Section~\ref{sec:computational_issues} we will show that these matrices {do not need to be explicitly built}, differently from $\mathcal{L}_k$. 
The next step in our construction is to use the information from all the $\mathbb{L}_k$ to build a matrix that accounts for walks of any length $k = 1,2, \ldots$ in the graph. To ensure the convergence of the resulting matrix and to align with the intuition behind the construction of walk-based centralities~\cite{MR2736969}, we present a definition that includes a monotonically decreasing sequence of coefficients $\{c_k\}_{k \geq 0}$ to weight the contributions of walks of different lengths. 
\begin{definition}\label{def:transformedlaplacian}
For any {sequence $\{c_k\}_{k\geq 0}$ that is monotonically non-increasing for $k \geq 1$,} %
and such that $\sum_{k=0}^{\infty} c_k A^k$ is convergent, the {$k$-walk transformed Laplacian} $\mathbb{L}(\mathbf{c})$ is defined as:
\[
\mathbb{L}(\mathbf{c}) = \operatorname{diag}\left( \sum_{k=0}^{\infty} c_k A^k \mathbf{1} \right) -  \sum_{k=0}^{\infty} c_k A^k = \sum_{k=1}^{\infty} c_k \mathbb{L}_k.%
\]
\end{definition}

It is easy to verify that, for $\mathbf{c} = (0, 1,0,0,\ldots)$, we recover $\mathbb{L}(\mathbf{c}) = L$. 

We note that, differently from the $k$-path transformed Laplacian $\mathcal{L}$ in~\eqref{eq:kpathlaplacian-global}, the $k$-walk transformed Laplacian $\mathbb{L}(\mathbf{c})$ does not reduce to a finite sum.  
However, whenever the coefficients $\{c_k\}_{k \geq 0}$ in Definition~\ref{def:transformedlaplacian} are chosen so that there exists a function $f$ for which $f(x) = \sum_{k \geq 0} c_k x^k$ on an open set $\Omega \subset \mathbb{C}$ containing the spectrum of $A$~\cite[Definition~1.1]{MR2396439}, then 
\begin{equation}\label{eq:f-laplacian}
\mathbb{L}(f) = \operatorname{diag}(f(A)\mathbf{1}) - f(A).
\end{equation}
In the following, we will restrict ourselves to sequences of coefficients that allow for the latter representation of the $k$-walk transfer Laplacian.

We postpone the discussion on the computational issues related to the evaluation of $\mathbb{L}(f)$ to Section~\ref{sec:computational_issues}. The remainder of this section is devoted to theoretical properties of $\mathbb{L}(f)$ and to one generalization.

\subsection{Analogies and relationship with the standard Laplacian}
We begin by pointing out that we can refine the observation made in Section~\ref{sec:laplacian_matrix_and_diffusion_equation} about the properties of the diffusion equation~\eqref{eq:diffusion_on_G}, expressed through the Laplacian matrix~$L$, and highlight the connections between the classical construction and the framework introduced here. 

The following proposition summarizes classical results that we used in Section~\ref{sec:notation_and_bg} and that we will show extend to $\mathbb{L}(f)$.
\begin{proposition}\label{pro:the-classical-L-gives-us}
For an unweighted, simple graph $G = (V, E)$, the Laplacian matrix $L$ satisfies the following.
\begin{itemize}
\item $L$ is symmetric;
\item $L\bone=\mathbf{0}$; and
\item $L$ is an M-matrix.
\end{itemize}
\end{proposition}
We have already discussed within Section~\ref{sec:laplacian_matrix_and_diffusion_equation} the relationship between the diffusion equation and the fact that $L$ is an M-matrix. The following result shows that, under suitable assumptions on $f$, $\mathbb{L}(f)$ is also an M-matrix, and thus it is possible to define a diffusion equation on the graph extending~\eqref{eq:diffusion_on_G} based on the idea of walks as 
\begin{equation}\label{eq:diffusion_on_G_with_walk}
\begin{cases}
\displaystyle \frac{\mathrm{d} \mathbf{p}(t)}{\mathrm{d} t} = - \mathbf{p}(t) \mathbb{L}(f), & t \in [0, t^*], \\
\mathbf{p}(0) = \mathbf{p}_0.
\end{cases}
\end{equation}
\begin{proposition}\label{pro:still-an-mmatrix}
Let $G = (V,E)$ be a simple connected graph and let $f$ be a function defined on the spectrum of its adjacency matrix $A\in\mathbb{R}^{n \times n}$; cf. \cite[Definition~1.1]{MR2396439}. Then the $k$-walk transformed Laplacian $\mathbb{L}(f)$ in~\eqref{eq:f-laplacian} is such that
\begin{itemize}
    \item $\mathbb{L}(f)$ is symmetric;
    \item $\mathbb{L}(f) \mathbf{1} = \mathbf{0}$; and  
    \item If $f(A) \geq 0$, then $\mathbb{L}(f)$ is an M-matrix. %
\end{itemize}
\end{proposition}

{\begin{proof}
The matrices $f(A)$ and $\mathbb{L}(f)$ are symmetric by construction, since $A^\top=A$. The fact that $\mathbb{L}(f) \mathbf{1} = \mathbf{0}$ follows trivially from~\eqref{eq:f-laplacian}.  
Suppose now that $f(A)\geq 0$. Then, by Lemma~\ref{lem:being_an_m_matrix}, $\mathbb{L}(f)$ is a singular M-matrix. 
\end{proof}}

For the purposes of this work, Proposition~\ref{pro:still-an-mmatrix} constrains our choice of $f$ to those functions that preserve nonnegativity. Characterizing these is a well studied problem and several results are available in the literature; see, for example, \cite{MR2735981,MR2399570} and references therein. If we restrict our attention to functions $f$ that generate admissible centrality indices, such as those belonging to class \(\mathcal{P}\) as introduced in \cite{MR3354998}\footnote{The class $\mathcal{P}$ contains all analytic functions that can be expressed as sums of power series with strictly positive coefficients on some open neighborhood of $0$.}, then the induced $\mathbb{L}(f)$ is an M-matrix. Furthermore, selecting $f\in\mathcal{P}$ allows us to naturally interpret the role of \(f(A)\) within the \eqref{eq:f-laplacian} in terms of centrality and to provide insight in the diffusion process modeled by $\mathbb{L}(f)$. 
Indeed, recall that the vector of total communicability centralities~\cite{totcomm} of nodes is defined, for a function $f$ defined on the spectrum of $A$, as:
\[
\mathbf{t}(f) = f(A)\mathbf{1}.
\]
When $f(x) = x$, it immediately follows that $\mathbf{t}(f) = (A\bone)$, and thus that $\mathbb{L}(f) = L$. However, for other choices of $f\in\mathcal{P}$, the introduced $k$-walk transformed Laplacian is simply generalizing $L$ by replacing degree centrality $A\bone$ with the total communicability $\mathbf{t}$ of nodes. With this interpretation, we can deduce that the edge centric diffusion equation~\eqref{eq:diffusion_on_G_with_walk} induced by $\mathbb{L}(f)$ is one where a walker tends to leave high total communicability nodes faster than low total communicability ones. Clearly, when $f(x) = x$, \eqref{eq:diffusion_on_G_with_walk} boils down to~\eqref{eq:diffusion_on_G}

\subsection{Backtrack-downweighted \texorpdfstring{$k$-walk}{k-walk} transformed Laplacian}

As described in the introduction, we are interested in defining a generalization of the Laplacian operator that accounts for walks on graphs. 
This we have partially achieved already in the previous section. Our next goal is to further generalize the definition in order to allow the flexibility to downweight (or possibly eliminate) certain walks in our count. 

\begin{definition}\label{def:NBT}
Given a graph $G=(V,E)$ without loops, a walk on $G$ is said to be \textit{backtracking} if it contains at least one instance of the form $i\to j \to i$, for $i,j\in V$, \textit{nonbacktracking (NBT)} otherwise. 
\end{definition}

We now introduce matrices $p_k(A)\in\mathbb{R}^{n\times n}$, for $k\in\mathbb{N}$, which
generalize to the NBT setting the idea that powers of the adjacency matrix encode the number of walks of a certain length in the graph; see Proposition~\ref{pro:power-of-adjacency}. 
These matrices are entrywise defined as
\[
(p_k(A))_{ij} = |\{\text{NBT walks of length } k \text{ from } i \text{ to } j \}|, \qquad k\in\mathbb{N}, \, i,j\in V
\]
and are symmetric by construction.  
It can be shown~\cite{MR271401}, \cite[Lemma 1]{MR1399606} that the matrices $p_k(A)$ satisfy a three terms recurrence. 
Setting $p_0(A) = I$, $p_1(A) = A$ and $p_2(A) = A^2 - D$, where $D$ is the degree matrix, for all $k\geq 3$ it holds 
\begin{equation}\label{eq:Br}
    p_{k}(A) = A\, p_{k-1}(A) + (I - D)\, p_{k-2}(A).
\end{equation}

This result was generalized in \cite{MR4296758}, where backtracking was downweighted by a factor $\theta = 1-\mu$, with $\mu\in[0,1]$, rather than completely eliminated. 

Following \cite{MR4296758}, we will let $q_{k}(A)$ denote the \textit{backtrack-downweighted walk (BTDW)} count matrix, where, for brevity, 
the dependence on $\mu$ is not made explicit. 
More precisely, 
$(q_k(A))_{ij}$ is the weighted number of distinct walks
of length $k$ from node $i$ to node $j$ where the weights are assigned according to the following; 
for each walk,  $i=i_0, i_1, i_2, \ldots, i_k=j$,
every occurrence of a backtracking step
($i_{\ell} = i_{\ell+2}$)
incurs a downweighting by a factor $1-\mu$. 
Then, it can be shown that, setting $q_0(A) = I$, 
$q_1(A) = A$ and $q_2(A) = A^2 - \mu D$, the BTDW matrices satisfy a three terms recurrence
\begin{equation}\label{eq:qrec}
q_{k+1}(A) = A q_k(A) + \mu \, \left( \mu I - D\right)q_{k-1}(A).
\end{equation}
Equation~\eqref{eq:qrec} specifies \cite[Theorem 3.1]{MR4296758} to the case of undirected graphs.

It is easily verified that for \(\mu = 0\), the relation~\eqref{eq:qrec} reduces to the counting formula in Proposition~\ref{pro:power-of-adjacency}, while for \(\mu = 1\) it recovers~\eqref{eq:Br}. We are now in the position to construct the desired extension of the Laplacian matrix.

\subsubsection{Building the BTDW Laplacian}\label{sec:walk_nbt_laplacians}
The definition of $k$-walk transformed Laplacian~\eqref{eq:f-laplacian} we have put forward considers all traversals on the graph, in contrast to the $k$-path transformed Laplacian~\eqref{eq:kpathlaplacian-global}, which only allows paths. A middle ground between these two approaches, which specializes the walk count while not becoming too computationally taxing, is to consider downweigthings of backtracking walks. 

\begin{definition}\label{def:BTDW_k-Laplacian}
    In the above notation, for $\mu\in[0,1]$, the \textit{backtrack-downweighted (BTDW) $k$-walk Laplacian} is defined as 
    \[
    \mathbb{L}_{k,\mu} = \operatorname{diag}(q_k(A)\bone) - q_k(A).
    \]
\end{definition}

There are two substantial differences between the $k$-path and the BTWD $k$-walk Laplacians, besides the type of traversals being considered. The first difference is that there exist both a maximal and a minimal $k$-path Laplacian. Specifically, $\mathcal{L}_k$ is the zero matrix for every $k\geq \operatorname{diam}(G)$, while $\mathbb{L}_{k,\mu}$ is non-zero for every $k$, as long as $\mu\neq 1$ or $G$ is not a tree. 
The second difference is that, in the construction of $\mathbb{L}_{k,\mu}$, we consider weighted ``adjacency matrices" $q_k(A)$  that account for BTDW walks, whereas the ``adjacency matrices" appearing in the construction of the $k$-path Laplacian are unweighted, as their entries \cite[Equation 3.1]{RandomHopper} equal one if \textit{at least one} path of the given length exists between the corresponding two nodes.

To build BTDW $k$-walk transfer Laplacian, we need to characterize the convergence of the series $\sum_{k = 0}^{\infty} c_k q_k(A)$ and {the function it converges to}. To this end, we define 
for any integer $s \geq 0$,
\begin{equation}\label{eq:fs}
f_s(x) = \sum_{k=0}^{\infty}  c_{s+k} x^k,
\end{equation}
and prove a version of \cite[Theorem~5.4]{MR3842589} for the undirected case.

\begin{proposition}\label{pro:whe-we-converge}
Assume that the series \eqref{eq:fs} for $s=0$ has finite or infinite radius of convergence $\mathcal{R}$, and let $\mu \in [0,1]$. Then 
\[
\phi(A,\mathbf{c}) = \sum_{k = 0}^{\infty} c_k q_k(A)
\]
converges for $\rho(Z) < \mathcal{R}$, where $\rho(Z)$ {is} the spectral radius of
\begin{equation}     \label{eq:Zdef}
 Z = 
\begin{bmatrix}
    O & I \\
      \mu(\mu I - D) 
         & 
         A
\end{bmatrix}.
\end{equation}
Furthermore,
\[
\phi(A,\mathbf{c}) = 
\begin{bmatrix}
O & I 
\end{bmatrix} f_2(Z) \begin{bmatrix}
q_1(A) \\
q_2(A)
\end{bmatrix} + c_1 A + c_0 I
=
\begin{bmatrix}
I & O 
\end{bmatrix} f_1(Z) \begin{bmatrix}
q_1(A) \\
q_2(A)
\end{bmatrix} + c_0 I.
\]
\end{proposition}

\begin{proof}
We begin by expanding the defining series
\begin{equation}\label{eq:series_to_be_shifted}
\phi(A,\mathbf{c})
    = \sum_{k=0}^{\infty} c_k  q_k(A)
    = c_0 I + c_1  A + \sum_{k=2}^{\infty} c_k  q_k(A).
\end{equation}
It is easy to verify  
by induction that 
\[
Z^k\begin{bmatrix}q_1(A)\\q_2(A)\end{bmatrix} = \begin{bmatrix}q_{k+1}(A)\\q_{k+2}(A)\end{bmatrix}.
\]
This, together with \eqref{eq:series_to_be_shifted}, yields:
\[
\sum_{k=2}^{\infty} c_k q_k(A)
= \sum_{k=0}^{\infty} c_{k+2} q_{k+2}(A)
=
\begin{bmatrix}O & I\end{bmatrix}
\Big(\sum_{k=0}^{\infty}c_{k+2}Z^k\Big)
\begin{bmatrix}q_1(A)\\ q_2(A)\end{bmatrix},
\]
and thus the first expression for $\phi(A,\mathbf{c})$ is readily recovered using~\eqref{eq:fs} with $s=2$. A similar derivation allows to obtain the second expression in our statement.
Finally, since \(f_0\) has radius of convergence \(\mathcal R\), analytic functional calculus gives that \(f_2( Z)\) converges whenever 
$\rho(Z)<\mathcal R$, which establishes both convergence and the closed-form expressions.
\end{proof}

\begin{corollary}\label{cor:when-we-converge}
For the case $\mu = 1$ the identity in Proposition~\ref{pro:whe-we-converge} reduces to:
\[
\phi(A,\mathbf{c}) = \sum_{k=0}^{\infty}  c_k  p_k(A) = \begin{bmatrix}
I & O 
\end{bmatrix} f_1( Z) \begin{bmatrix}
A \\
A^2 - D
\end{bmatrix} + c_0 I.
\]
\end{corollary}

\begin{remark}
If $\mathcal{R}<\infty$ and $\rho(Z) > \mathcal{R}$ in either Proposition~\ref{pro:whe-we-converge} or Corollary~\ref{cor:when-we-converge}, one can introduce a parameter $0 < \alpha < 1$ to rescale the series in~\eqref{eq:fs} as $f_s(\alpha\, x)$. Such $\alpha$ will thus need to be chosen so that $\alpha \rho(Z) < \mathcal{R}$. 
\end{remark}

We now mirror the construction in Definition~\ref{def:transformedlaplacian} and introduce an operator that includes BTDW walks of all possible lengths. 
\begin{definition}\label{def:BDTW-Laplacian}
For any  {sequence $\{c_k\}_{k\geq 0}$ that is monotonically non-increasing for $k \geq 1$ and} such that $\sum_{k = 0}^{\infty} c_k Z^k$ is convergent, and for any $\mu \in [0,1]$ we define the \textit{BTDW $k$-walk transformed Laplacian} as
\begin{equation}\label{eq:weighted_laplacian}
\mathbb{L}_{\mu}(\mathbf{c}) =  \operatorname{diag}\left( \sum_{k \geq 0} c_k  q_k(A) \mathbf{1} \right) - \sum_{k \geq 0} c_k  q_k(A)  = \sum_{k = 1}^{\infty} c_k \mathbb{L}_{k,\mu}.
\end{equation}
\end{definition}

The following result, which is the {analogue} of Proposition~\ref{pro:still-an-mmatrix} for $\mathbb{L}_\mu(\mathbf{c})$, describes the characterization we need in order to formulate a diffusion equation employing the newly introduced operator.
\begin{proposition}\label{pro:still-again-an-mmatrix}
Let $G = (V,E)$ be a simple connected graph and let $f_s$, $s=0,1$, be a function defined on the spectrum of $Z \in\mathbb{R}^{2n \times 2n}$ for $Z$ as in Proposition~\ref{pro:whe-we-converge}. Then the BTDW $k$-walk transformed Laplacian $\mathbb{L}_\mu(\mathbf{c})$, $\mu \in (0,1]$, in~\eqref{eq:weighted_laplacian} is such that
\begin{itemize}
    \item $\mathbb{L}_{\mu}(\mathbf{c})$ is symmetric;
    \item $\mathbb{L}_{\mu}(\mathbf{c}) \mathbf{1} = \mathbf{0}$; and  
    \item If $\phi(A,\mathbf{c})\geq 0$, then $\mathbb{L}_{\mu}(\mathbf{c})$ is an M-matrix. %
\end{itemize}
\end{proposition}
{\begin{proof}
Symmetry holds true by construction. 
Moreover,
$\mathbb{L}_{\mu}(\mathbf{c})\mathbf{1}
=\operatorname{diag}(\phi(A,\mathbf{c})\mathbf{1})\mathbf{1}
-\phi(A,\mathbf{c})\mathbf{1}
=\mathbf{0}$,
which proves the second claim. Finally, suppose that $\phi(A,\mathbf{c})\geq0$. By taking $C=\phi(A,\mathbf{c})$ in Lemma~\ref{lem:being_an_m_matrix}, it immediately follows that 
$\mathbb{L}_{\mu}(\mathbf{c})
=\operatorname{diag}(\phi(A,\mathbf{c})\mathbf{1})-\phi(A,\mathbf{c})$
is a singular M-matrix.
\end{proof}}

We point out that any function $f = f_0\in\mathcal{P}$ ensures nonnegativity of $\phi(A,\mathbf{c})$ since
\[
\begin{bmatrix}     \delta_{s,1}I & \,\delta_{s,2}I \end{bmatrix}f_s(Z) \begin{bmatrix}     q_1(A) \\ q_2(A) \end{bmatrix} = 
\sum_{k=0}^\infty c_{k+s}q_k(A) \geq 0
\]
where $\delta_{i,j}$ is the Kronecker delta. 
The added value of selecting  $f\in\mathcal{P}$ is that any of these functions would allow the interpretation of $\mathbb{L}_\mu(\mathbf{c})$ in terms of centralities~\cite{MR3801714}. Henceforth, for all $\mathbf{c}$ obtained from $f \in \mathcal{P}$ we will write $\mathbb{L}_\mu(f)$ to denote $\mathbb{L}_\mu(\mathbf{c})$. 

By means of Definition~\ref{def:BDTW-Laplacian} we can extend the diffusion equation~\eqref{eq:diffusion_on_G_with_walk} to
\begin{equation}\label{eq:diffusion_on_G_with_downweighted_walk}
\begin{cases}
\displaystyle \frac{\mathrm{d} \mathbf{p}(t)}{\mathrm{d} t} = - \mathbf{p}(t) \mathbb{L}_\mu(f), & t \in [0, t^*], \\
\mathbf{p}(0) = \mathbf{p}_0,
\end{cases} \qquad \mu \in [0,1], 
\end{equation}
which reduces exactly to~\eqref{eq:diffusion_on_G_with_walk} for $\mu = 0$. 
In fact, it is easy to see that, when $\mu = 0$ and backtracking is not downweighted, $\mathbb{L}_0(\mathbf{c}) = \mathbb{L}(\mathbf{c})$ and moreover Proposition~\ref{pro:still-again-an-mmatrix} reduces to Proposition~\ref{pro:still-an-mmatrix}.
For $\mu = 1$ we get instead the NBT version of the $k$-walk transformed Laplacian. In the following, we provide two explicit examples of NBT Laplacians built using the two matrix functions  most widely used in network science applications; the resolvent and the exponential.

\paragraph{The resolvent NBT \texorpdfstring{$k$-walk}{k walk} transformed Laplacian.} We first consider the case of the \emph{resolvent}, when $\mathbf{c} = (1,\alpha,\alpha^2,\ldots)$ with $\alpha \rho(Z) < 1$, which is a function in the $\mathcal{P}$ class. Then, by Corollary~\ref{cor:when-we-converge}, 
\begin{equation}\label{eq:deformed}
\phi(A,\mathbf{c}) [ I - \alpha A - \alpha^2(I-D) ]  = (1-\alpha^2) I.
\end{equation}
Setting $\mathcal{A}(\alpha) =  I - \alpha A - \alpha^2(I-D)$, the corresponding NBT $k$-walk transformed Laplacian is
\begin{equation}\label{eq:katz-based-walk-laplacian}
\mathbb{L}_{1}\left( (1-\alpha x)^{-1} \right) = (1-\alpha^2) \left( \operatorname{diag}\left( \mathcal{A}(\alpha)^{-1} \mathbf{1} \right)  - \mathcal{A}(\alpha)^{-1} \right),
\end{equation}
which correspond to the matrix having on the diagonal the NBT Katz centrality vector~\cite{MR3801714}.
\begin{remark}
The matrix $\mathcal{A}(\alpha)$ is the \emph{deformed graph Laplacian} which has been used to investigate consensus problems leveraging the spectral theory of quadratic eigenvalue problems~\cite{MR3092654}; furthermore, for undirected graphs, its inverse $\mathcal{A}(\alpha)^{-1}$ is known as the Bethe Hessian~\cite{NIPS2014_63923f49} and has been employed in spectral clustering tasks.    
\end{remark}

In general, \(\mathcal{A}(\alpha)\) is symmetric but indefinite~\cite[p.~3050]{MR3092654}.  However, by choosing \(\alpha\) appropriately, positive definiteness can be recovered; in particular, one may select \(\alpha\) so that \(\mathcal{A}(\alpha)\) becomes an M-matrix. {The proposition below characterizes such choices of $\alpha$, summarizing results from \cite{MR3769701}.}
\begin{proposition}\label{prop:CGworks}
Let $G$ be a connected simple graph {with average degree $> 2$}. 
Then, {for every $0<\alpha <1/\rho(Z)$, the deformed graph Laplacian $\mathcal{A}(\alpha)$ is an M-matrix.}
\end{proposition}
{\begin{remark} The hypothesis of a connected simple graph having average degree at least two rewrites as $2m > 2n$, where $m$ is the number of undirected edges in the graph. Since normally $m\approx \log(n)$ in real-life networks, this hypothesis is always satisfied in practice.\end{remark}}

{\begin{remark}
    Equation~\ref{eq:deformed} and Proposition~\ref{prop:CGworks} readily extend to the general case of $\mu\in[0,1]$, by adapting the results in~\cite[Theorems~4.1 and 6.1]{MR4296758} and employing again Lemma~\ref{lem:being_an_m_matrix} to prove the property of being an M-matrix.
\end{remark}}

\paragraph{The exponential NBT \texorpdfstring{$k$-walk}{k walk} transformed Laplacian.} Another example of choosing the specific weights for~\eqref{eq:weighted_laplacian} is that of the exponential $f_0(x) = e^x$ in~\eqref{eq:fs}, which corresponds to
\[
f_1(x) = \sum_{k = 1}^{\infty}  \frac{1}{(k+1)!} x^k = \frac{e^x-1}{x} = \varphi_1(x),
\]
i.e., to the first $\varphi$-function from the theory of exponential integrators~\cite{MR2652783}
\[
\varphi_0(x) = e^x, \quad \varphi_s(x) = \frac{1}{(s-1)!} \int_{0}^{1} e^{(1-\theta)x} \theta^{s-1}\,\mathrm{d}\theta, \qquad s=1,2,\ldots, \qquad y \in \mathbb{C}.
\]
These functions are {entire} and as such admit an infinite radius of convergence. The resulting exponential NBT $k$-walk transformed Laplacian is thus given by
\begin{equation}\label{eq:exp-based-laplacian}
\begin{split}
\mathbb{L}_{1}(\varphi_1) = & \; \operatorname{diag}\left( \begin{bmatrix}
I & O 
\end{bmatrix} \varphi_1(Z) \begin{bmatrix}
\mathbf{d} \\
A^2\mathbf{1} - \mathbf{d}
\end{bmatrix}  \right) -  \begin{bmatrix}
I & O 
\end{bmatrix} \varphi_1(Z) \begin{bmatrix}
A \\
A^2 - D
\end{bmatrix},
\end{split}
\end{equation}
where $\mathbf{d} = A\bone = D\bone$. 

\section{Computational considerations}\label{sec:computational_issues}

In this section we detail different computational features of the newly introduced Laplacian operators $\mathbb{L}_\mu(\mathbf{c})$, making a direct comparison with their path-based counterpart $\mathcal{L}$ from~\cite{MR2900722,MR3834211}. We also describe the algorithms implemented to perform the numerical experiments in the next Section~\ref{sec:numerical_experiments}.

The construction of $k$-path Laplacians $\mathcal{L}_k$ in~\eqref{eq:kpathlaplacian} requires the computation of the geodesic distance $\mathrm{d}(\cdot,\cdot)$ for every pair of nodes in the underlying network. This is one of the major disadvantages related to their use. In a directed weighted graph with positive or negative edge weights and no negative cycles this can be solved with the Floyd–Warshall algorithm~\cite{Floyd,Bernard,Warshall} that has a worst-case performance of $\Theta(n^3)$. If the graph is also sparse then the cost can be brought down to $O(n^2\log(n)+2mn)$ with Johnson's algorithm~\cite{MR0449710}{, where $m$ is the number of edges in the graph}. Both require storage of $n^2$ elements. It is also important to observe that, even in situations where only matrix-vector products involving $\mathcal{L}$ are needed, the computational cost is not reduced. Indeed, one still requires the computation of the entire distance matrix. 
The newly introduced Laplacians $\mathbb{L}_\mu(f)$, for $\mu\in[0,1]$ and $f\in\mathcal{P}$, on the other hand, allow the implementation and use of significantly lower computational cost algorithms. Furthermore, even though the matrices are dense, they {do not need to be} stored. 
Therefore, they share the same advantages as top-down approaches based on the computation of matrix-vector products with the standard Laplacian matrix~\cite{MR4130854,MR4412292}. Below we briefly address some strategies for tackling these computations efficiently.

\subsection{Walk-based Laplacians}\label{sec:krylov_is_the_way}

If we compare the expression for $\mathcal{L}$~\eqref{eq:kpathlaplacian-global} with the one for $\mathbb{L}_\mu(f)$~\eqref{eq:weighted_laplacian}, we can observe that, from a computational standpoint, the latter allows to compute matrix-vector products without ever assembling the whole matrix $\mathbb{L}_\mu(f)$. Krylov methods~\cite{MR3095912,MR2923559} provide an efficient approach for approximating objects of the form $f(A)\mathbf{v}$, as we need to do for~\eqref{eq:weighted_laplacian}. 
In the simplest case of a polynomial method, given a matrix $A \in \mathbb{R}^{n \times n}$ and a vector $\mathbf{v} \in \mathbb{R}^n$, we first construct the Krylov subspace of order $k$, defined as 
$ \mathcal{K}_k(A, \mathbf{v}) = \text{span}\{\mathbf{v}, A\mathbf{v}, A^2\mathbf{v}, \ldots, A^{k-1}\mathbf{v}\},$
by generating an orthonormal basis $\{\mathbf{q}_1, \mathbf{q}_2, \ldots, \mathbf{q}_k\}$ for it using either the Arnoldi (for $A\neq A^\top$, \cite[Algorithm~6.1]{MR1990645}) or Lanczos (for $A= A^\top$, \cite[Algorithm~6.15]{MR1990645}) process. The basis is stored in a rectangular matrix $Q_k = [\mathbf{q}_1, \mathbf{q}_2, \ldots, \mathbf{q}_k]$ which is then used to form the Hessenberg (resp., tridiagonal for $A= A^\top$), matrix $H_k = Q_k^\top A Q_k \in \mathbb{R}^{k \times k}$. The matrix $H_k$ represents the projection of $A$ onto $\mathcal{K}_k(A, \mathbf{v})$.
With this, one can then cheaply compute\footnote{{
The computation of $f(H_k)$ can be performed, e.g., via the Schur-Parlett method~\cite[\S 6.4]{MR2396439}, or, for particular choices of $f$, by means of specialized methods. We refer the interested reader to~\cite{MR2396439} for more details. 
}} the matrix function $f(H_k)$, since $k \ll n$, and then approximate $f(A)\mathbf{v}$ as $f(A)\mathbf{v} \approx  \|\mathbf{v}\| Q_k f(H_k) \mathbf{e}_1$, where we have used the fact that $Q_k^\top \mathbf{v} =  \|\mathbf{v}\| \mathbf{e}_1$.
Krylov subspace methods provide accurate approximations to $f(A)\mathbf{v}$ without the need to directly compute $f(A)$ for large and sparse matrices, an operation that would normally require $O(n^2)$ storage and $O(n^3)$ computation. 
In some cases it may be convenient to move from polynomial methods to rational ones~\cite{MR3095912,MR4082482}. In such instances, the cost to augment the dimension of the Krylov space increases, since we must move from the simple computation of matrix-vector products to the solution of a linear system at each space-expansion step. However, this usually has the advantage of significantly reducing the dimension of the space to be generated. Moreover, these methods leverage linear cost algorithms for the efficient solution of sparse linear systems; see Section~\ref{sec:prob_estimation} below. 
\subsubsection{Simulating discrete-time diffusion}\label{sec:markov}
Using the same procedure, we can also compute the action of the probability transition matrix $P$ of a Markov chain associated with the Laplacians in 
Definition~\ref{def:BDTW-Laplacian} for $\mu\in[0,1]$\footnote{Recall that, when $\mu = 0$ we recover Definition~\ref{def:transformedlaplacian}.} as 
\begin{equation}\label{eq:let_there_be_markov}
     P = I - {\mathbb{D}}^{-1}\mathbb{L}_\mu(f), 
     \qquad 
     {\mathbb{D}} = \operatorname{diag}(\mathbb{L}_\mu(f)).
\end{equation}
Constructing the diagonal matrix \({\mathbb{D}}\) only requires a single evaluation of the underlying matrix function; 
afterward, the same routine used to apply the chosen walk-based Laplacian may be reused efficiently.
\begin{remark}
This is not the case for the fractional Laplacian $L^\gamma$, for $\gamma \in (0,1)$~\cite{MR4130854,Riascos1,MR4412292} or for the $k$-path transformed Laplacian $\mathcal{L}$~\eqref{eq:kpathlaplacian-global}. For the former, computing the diagonal matrix $D{_\gamma} = \operatorname{diag}(L^\gamma\bone)$ needed to define $P = I - D{_\gamma}^{-1}L^\gamma$ requires $n$ matrix function evaluations; for the latter, one has to actually compute the whole matrix $\mathcal{L}$, i.e., compute the distance matrix associated with the graph.
\end{remark}

The matrix $P$ in~\eqref{eq:let_there_be_markov} describes a random walk on the graph in which the transition probability between nodes is  inversely weighted with the total communicability measure induced by the selected function $f\in\mathcal{P}$.  
The evolution of a probability distribution then takes the form
\[
\mathbf{p}_{k+1} = \mathbf{p}_k P, 
\qquad 
\text{given } \mathbf{p}_0 \text{ with } \mathbf{p}_0 \mathbf{1} = 1.
\]

We emphasize that, while the distributions \(\mathbf{p}(t)\) in \eqref{eq:diffusion_on_G}, 
\eqref{eq:diffusion_on_G_with_walk}, and \eqref{eq:diffusion_on_G_with_downweighted_walk} evolve continuously under a Laplacian operator, 
the discrete-time Markov chain defined by \(P\) can be interpreted as the natural discrete analogue of the continuous diffusion equations.

\subsubsection{Limitations due to floating point arithmetic}

For both Definitions~\ref{def:transformedlaplacian} and~\ref{def:BDTW-Laplacian}, a note of caution is required regarding the magnitude of the entries of the matrices considered. 
When the function $f$ exhibits rapid growth with respect to its argument, e.g., when $f$ is the exponential function, the entries of $\mathbb{L}_\mu(f)$, $\mu\in[0,1]$ may grow so large to approach the overflow threshold of double-precision arithmetic. 
This effect is particularly evident on the diagonal elements, as it was previously observed in~\cite{MR4340667}. In such cases, although Propositions~\ref{pro:still-an-mmatrix} and~\ref{pro:still-again-an-mmatrix} remain valid, the resulting matrices may be numerically undefined or lead to severe ill-conditioning in floating-point computation. 
A practical remedy to this issue is to rescale the spectrum by introducing an ``inverse temperature'' within the function; for example, instead of using $f(x) = \exp(x)$, one may consider $f(x) = \exp(\beta x)$ with $\beta \propto 1/\rho(A)$.

\subsection{The resolvent NBT $k$-walk transformed Laplacian: $\mathbb{L}_1((1-\alpha x)^{-1})$}\label{sec:solving_katz}

Computation with the resolvent NBT $k$-walk transformed Laplacian requires the evaluation of matrix-vector products with $\mathbb{L}_1(f)$ from~\eqref{eq:katz-based-walk-laplacian}, where $f(x) = (1-\alpha x)^{-1}$ for suitable values of $\alpha>0$. Each of these amounts to solving a linear system defined in terms of the deformed Laplacian $\mathcal{A}(\alpha)$. If $\alpha$ is selected within the bounds from Proposition~\ref{prop:CGworks}, then this solution can be computed using the Conjugate Gradient algorithm. 
In general, this method requires a suitable preconditioner to ensure fast convergence. Since $\mathcal{A}(\alpha)$ is an M-matrix, we can utilise Algebraic MultiGrid methods (AMG), which are known to be especially effective for such operators. This is because their spectral behavior and smoothing properties align well with the coarse space construction and relaxation strategies used in AMG. 
For our task, we evaluate the usage of an AMG from the \texttt{PSCToolkit} library~\cite{DAMBRA2023100463}. Specifically, we employ an aggregation-based multigrid with the decoupled Van\v{e}k, Mandel, and Brezina aggregation~\cite{MR1393006} in a V-cycle, using one sweep of $\ell_1$-Jacobi as smoother~\cite{MR2861652} and 30 smoother sweeps as coarse grid solver. 
We will show in Section~\ref{sec:numerical_experiments} that this choice is computationally efficient and that it can also be used to perform the computation transparently on one or more GPU accelerators.

\subsection{Exponential NBT $k$-walk transformed Laplacian: $\mathbb{L}_1(exp)$}

The computation of matrix-vector products with the NBT $k$-walk framework Laplacian $\mathbb{L}_{1}(f)$ from~\eqref{eq:exp-based-laplacian} requires the evaluation of $\varphi_1(x)$. 
As previously mentioned, this is a common task when using exponential integrators for the numerical approximation of ODEs~\cite{MR2652783}. 
As it generally happens for the computation of matrix functions, the available algorithms are divided between the ones working on small-dimensional matrices~\cite{MR2492291,MR2785959} and those based on Krylov-type methods~\cite{MR2923559,DEKA2023101302}. 
Here, we use the algorithm described and implemented in~\cite{MR2923559}, which exploits a polynomial Krylov method.

\subsection{Average Return Probability and its Randomized Estimates}\label{sec:prob_estimation}

The efficiency of a continuous time random walk can be studied by considering the \textit{return probabilities}, i.e., the probability $[P(t)]_{j,j}$ of having started from node $j$ at time~$0$ and being back in $j$ at time $t$ following any of the evolution equations 
\[
\begin{cases}
\displaystyle \frac{\mathrm{d} \mathbf{p}(t)}{\mathrm{d} t} = - \mathbf{p}(t) \mathfrak{L}, & t \in [0, t^*], \\
\mathbf{p}(0) = \mathbf{p}_0, 
\end{cases}
\]
for some matrix $\mathfrak{L}$. 
If the initial decay of return probabilities occurs rapidly, the information swiftly traverses from the initial node to the entire network. Conversely, a gradual decay implies that the information can only propagate through the network at a slower pace. The desirability of either scenario depends on the specific phenomenon we aim to describe. The \textit{average return probability} for the evolution equations above is then defined as 
\begin{equation}\label{eq:average_return_probabilities}
\begin{split}
\hat{p}_0(t) = &\; \frac{1}{n} \sum_{i=1}^{n} [P(t)]_{ii} = \frac{1}{n} \operatorname{tr}\left( \exp(- t \mathfrak{L} ) \right) %
\end{split}
\end{equation}
This quantity allows us to make a global statement on the system's effectiveness in diffusing the information, thus assessing its overall efficiency.
In our experiments we will consider $\mathfrak{L}\in\{L,\mathcal{L},\mathbb{L}_\mu(f)\}$; see Section~\ref{sec:numerical_experiments} below. 

The computation of the average return probabilities in~\eqref{eq:average_return_probabilities} requires the evaluation of the trace of a matrix exponential for different values of $t$, nominally, until the quantity has changed from the initial value~$1$ to the final value~$1/n$. In general, computing the trace of a matrix function exactly costs $O(n^3)$ computation and $O(n^2)$ storage. 
We have to consider the entire matrix $\exp(-t \mathfrak{L})$ in order to compute its trace, even if only temporarily. 
We note, however, that we can be satisfied with an approximation to $\hat{p}_0(t)$, since our goal is purely descriptive and we only need it to be able to compare the different network exploration strategies described in Section~\ref{sec:kwalkdiffusion}. 
Two convenient sets of tools for this task are represented by quadrature-based estimators~\cite{MR4530345} and stochastic trace estimators~\cite[Section~4]{RandNLA}.
           \begin{algorithm}[t]
                \caption{\NysTraceExp: randomized average return probability estimate\label{alg:trace_estimation}}
                \begin{algorithmic}[1]
                        \Require Laplacian matrix $\{ L,\mathbb{L}_\mu(f),\mathbb{L}(f) \} \ni \mathfrak{L} \in \mathbb{R}^{n \times n}$, number $M$ of matvecs, times $t \in [0,t^*]$.
                        \Ensure Estimate $\hat{p}(t) \approx \frac{1}{n}\operatorname{tr}\left(\exp(-t \mathfrak{L})\right)$, and error estimate $\hat{e}(t) \approx \left\vert  \hat{p}(t) - \frac{1}{n} \operatorname{tr}\left(\exp(-t \mathfrak{L})\right)\right\vert$ for all $t \in [0,t^*]$.
                        \State Draw iid isotropic $\boldsymbol{\omega}_1,\ldots,\boldsymbol{\omega}_M \in \mathbb{R}^n$
                        \State $\boldsymbol{\Omega} \leftarrow \begin{bmatrix} \boldsymbol{\omega}_1 & \dots & \boldsymbol{\omega}_M\end{bmatrix}$
                        \State Approximate the spectral radius $\rho(\mathfrak{L})$ of $\mathfrak{L}$, and build the poles $\{\xi_j\}_{j=1}^{k+1}$ of rational approximation of $\exp(-x)$ on the interval $[0,t^* \rho(\mathfrak{L})]$.\Comment{Use \texttt{AAA}~\cite{MR3805855}}
                        \State Build the basis $V_{k+1}$ of $\mathcal{Q}^{\text{block}}_{k+1}(L, \boldsymbol{\Omega}, q_k)$ and the block upper Hessenberg matrix pencil \( (\mathcal{H}_k, \mathcal{K}_k) \). \Comment{matvecs/solve} 
                        \For{$t \in [0,t^*]$}\Comment{Can be a \textbf{parfor}}
                        \State ${Y(t)} \leftarrow \frac{1}{n} V_k \exp\left(- t \hat{\mathcal{H}}_k \hat{\mathcal{K}}_k^{-1} \right) V_k^\top \mathbf{\Omega}$ \Comment{Use the \textit{reduced} pencil}
                        \State $\hat{p}(t),\hat{e}(t) \gets \NysTrace(Y(t),\boldsymbol{\Omega})$
                        \EndFor
                \end{algorithmic}
        \end{algorithm}
Here, we focus on the latter, and adapt \NysTrace from~\cite{XTRACE} to our setting, as outlined in~Alg.~\ref{alg:trace_estimation}. Since we need to compute the matrix exponential for a range of parameters \( t \in [0,t^*] \), two distinct regions of approximation arise; one dominated by the larger eigenvalues and another where accuracy on the smaller eigenvalues is essential. For the first region, a polynomial Krylov method would suffice; however, its accuracy diminishes when focusing on the smaller eigenvalues. To address this, we use a Block Rational Krylov method. Recall that this method defines a linear space of block vectors of size \( n \times M \), constructed with respect to the matrix \( L \in \mathbb{R}^{n \times n} \) and a starting block vector of maximal rank:
\[
\mathbf{\Omega} = [\boldsymbol{\omega}_1, \boldsymbol{\omega}_2, \dots, \boldsymbol{\omega}_M] \in \mathbb{R}^{n \times M}, \quad \operatorname{rank}(\mathbf{\Omega}) = M.
\]
To define the block rational Krylov space, we require a sequence of poles \( \xi_1, \xi_2, \dots, \xi_k \), which are complex numbers or infinity, all distinct from the eigenvalues of \( \mathfrak{L} \). The block rational Krylov space of order \( k+1 \), associated with \( \mathfrak{L}, \mathbf{\Omega}, \xi_j \), is defined as
\begin{equation}\label{eq:block-rational-Krylov}
    \mathcal{Q}^{\text{block}}_{k+1}\left( \mathfrak{L}, \mathbf{\Omega}, r_k \right) = r_k\left( \mathfrak{L} \right)^{-1} \left\{ \sum_{i=0}^{k} \left( \mathfrak{L} \right)^i \mathbf{\Omega} C_i \right\},
\end{equation}
where \( \{ C_i \}_{i=0}^{k} \) are matrices of size \( M \times M \), and
\[
r_k(z) = \prod_{\substack{j=1 \\ \xi_j \neq \infty}}^{k} (z - \xi_j)
\]
is the common denominator of the rational matrix-valued functions associated with the block rational Krylov space. The construction of the basis vectors requires the solution of linear systems with matrix $\mu_k \mathfrak{L} - \nu_k I$, and $\mu_k$ and $\nu_k\in\mathbb{C}$ obtained from the poles, in the case of Laplacians of the type \eqref{eq:f-laplacian} and \eqref{eq:weighted_laplacian}. However, we only have access to routines that allow us to perform the matrix-vector product (Section~\ref{sec:krylov_is_the_way}). This means that we cannot use a direct solution, but we can rely again on the use of a Krylov-type method for the solution of linear systems fixed to achieve a lower tolerance than that which we require for the entire approximation, e.g., we can use the GMRES method. To generate an orthogonal block basis for this space, we then employ the block rational Arnoldi method~\cite{MR4082482}, which produces
\[
V_{k+1} = [\mathbf{v}_1, \dots, \mathbf{v}_{k+1}] \in \mathbb{C}^{n \times (k+1)M}, \quad \mathbf{v}_j \in \mathbb{C}^{n \times M},
\]
satisfying the block rational Arnoldi decomposition:
\[
\mathfrak{L} V_{k+1} \mathcal{K}_k = V_{k+1} \mathcal{H}_k,
\]
where \( (\mathcal{H}_k, \mathcal{K}_k) \) is an unreduced block upper Hessenberg matrix pencil of size \( (k+1)M \times kM \). The block vectors in \( V_{k+1} \) span the space \( \mathcal{Q}^{\text{block}}_{k+1}\left( \mathfrak{L}, \mathbf{\Omega}, r_k \right) \).

To estimate the average return probabilities, we first sample \( M \) independent identically distributed isotropic vectors \( \{ \boldsymbol{\omega}_i \}_{i=1}^{M} \), and then construct the Krylov subspaces \( \mathcal{Q}^{\text{block}}_{k+1}\left( \mathfrak{L}, \mathbf{\Omega}, r_k \right) \). This involves computing the basis \( V_{k+1} \) and the unreduced block upper Hessenberg matrix pencil \( (\mathcal{H}_k, \mathcal{K}_k) \). For each \( t \in [0, t^*] \), we then compute
\[
Y(t) = \frac{1}{n}V_k \exp\left( -t \hat{\mathcal{H}}_k \hat{\mathcal{K}}_k^{-1} \right) V_k^\top \mathbf{\Omega},
\]
where the matrix-exponential is now efficiently computed on the reduced pencil---excluding the last block row of size \( M \)---\( (\hat{\mathcal{H}}_k, \hat{\mathcal{K}}_k) \). To complete the description, we need to determine the poles \( \{ \xi_j \}_{j=1}^{k} \) for the construction of~\eqref{eq:block-rational-Krylov}. Since we aim for an approximation of the matrix exponential that is accurate for all values of \( t \in [0, t^*] \), we select the poles using the \texttt{AAA} algorithm for rational approximation~\cite{MR3805855}. The support points for the approximation are sampled from the interval \( [0, t^* \rho(\mathfrak{L})] \).

Finally, we run the trace estimator \NysTrace on the matrix \( Y(t) \) and the block of vectors \( \mathbf{\Omega} \) for all $t \in [0,t^*]$. This step can be parallelized in shared memory, as the estimates are independent and only require sharing the Krylov subspace basis and the sampling vectors \( \mathbf{\Omega} \).

\section{Numerical experiments}\label{sec:numerical_experiments}

In this section we analyze the walk-based diffusion proposals introduced in Sections~\ref{sec:kwalkdiffusion} and~\ref{sec:walk_nbt_laplacians} both from the modelling and computational point of view. Specifically, in Sections~\ref{sec:small-examples} and~\ref{sec:averagereturn} we investigate the network-exploration properties of the different strategies, while in Section~\ref{sec:katzexperiment} we analyze the computational challenges posed by the models.

The tests are run on the graphs reported in Table~\ref{tab:complexnetworks} from the \texttt{SuiteSparse Matrix Collection}~\cite{suitesparse}. The code to reproduce the experiments is available on the \texttt{GitHub} repository \href{https://github.com/Cirdans-Home/walk-laplacian}{Cirdans-Home/walk-laplacian}.
\begin{table}[htbp]
\centering
\caption{Summary of networks in our dataset. For each we report the size ($n =|V|$) and the spectral radius of the matrix $Z$ for $\mu = 1$; see Proposition~\ref{pro:whe-we-converge}.}
\label{tab:complexnetworks}
\begin{tabular}{llrr}
\toprule
& Network name & $n$ & $\rho(Z)$ \\
\midrule
1 & \texttt{Arenas/PGPgiantcompo} & 10680 & 41.03 \\
2 & \texttt{Pajek/USpowerGrid} & 4941 & 6.23 \\
3 & \texttt{DIMACS10/belgium\_osm} & 1441295 & 2.58 \\
4 & \texttt{SNAP/ca-CondMat} & 21363 & 35.86 \\
5 & \texttt{SNAP/ca-HepPh} & 11204 & 243.75 \\
6 & \texttt{Pajek/dictionary28} & 24831 & 16.07 \\
7 & \texttt{Belcastro/human\_gene1} & 21853 & 3335.86 \\
8 & \texttt{Belcastro/human\_gene2} & 14020 & 2999.83 \\
9 & \texttt{SNAP/roadNet-CA} & 1957027 & 3.32 \\
10 & \texttt{SNAP/roadNet-PA} & 1087562 & 3.11 \\
11 & \texttt{SNAP/roadNet-TX} & 1351137 & 3.56 \\
12 & \texttt{Gleich/usroads-48} & 126146 & 2.76 \\
\bottomrule
\end{tabular}
\end{table}
The numerical experiments were conducted on the Toeplitz cluster at the \textit{Green Datacenter} of the University of Pisa. We employ the \texttt{gpu} nodes that are each equipped with a total of 256 CPUs, specifically AMD EPYC 7763 64-Core Processors \SI{3.5}{\giga\hertz}, with each processor providing 2 threads per core and 64 cores per socket, across 2 sockets. For the modeling and descriptive experiments in Sections~\ref{sec:small-examples} and~\ref{sec:averagereturn} we employed MATLAB \texttt{v.9.10.0.1602886 (R2021a)}. The experiments in Section~\ref{sec:katzexperiment} exploit the \texttt{PSCToolkit} library, and the availabe NVIDIA A40 GPUs (memory capacity of \SI{4.6}{\tera\byte} and a maximum power usage of \SI{300}{\watt}). For the latter, the NVIDIA driver \texttt{v.535.274.02} and CUDA \texttt{v.2.2} are used to manage these GPUs. The system software stack includes GCC \texttt{v.12.2.0}, OpenBLAS \texttt{v.0.3.26}, PSBLAS \texttt{v.3.9.0}, AMG4PSBLAS \texttt{v.1.2.0}.

\subsection{Descriptive examples}\label{sec:small-examples}

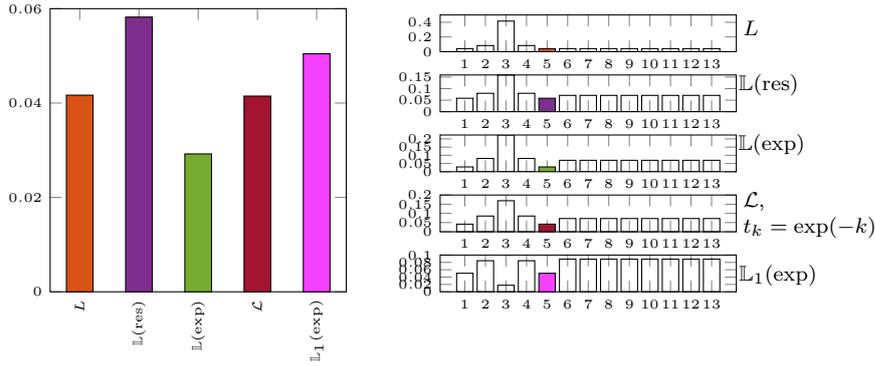
\begin{figure}[htbp]
    \centering
    \definecolor{laplacian}{HTML}{d95319}
\definecolor{katz}{HTML}{7e2f8e}
\definecolor{exp}{HTML}{77ac30}
\definecolor{estrada}{HTML}{a2142f}
\definecolor{nbtexp}{HTML}{f341ff}

\begin{tikzpicture}

\begin{axis}[%
width=0.328\columnwidth,
height=0.316\columnwidth,
at={(0\columnwidth,0\columnwidth)},
scale only axis,
xtick={1,2,3,4,5},
xticklabels={$L$,$\mathbb{L}(\operatorname{res})$,$\mathbb{L}(\exp)$,$\mathcal{L}$,$\mathbb{L}_1(\exp)$},
xticklabel style={rotate=90},
xmin=0.5,
xmax=5.5,
ymin=0,
ymax=0.06,
ticklabel style = {font=\tiny},
scaled y ticks=false,
yticklabel style={
        /pgf/number format/fixed,
        /pgf/number format/precision=2
},
axis background/.style={fill=white}
]
\addplot[ybar, draw=black, area legend, fill=laplacian] table[row sep=crcr] {%
1	0.0416666666666667\\
};
\addplot[ybar, draw=black, area legend, fill=katz] table[row sep=crcr] {%
2	0.0582171338987794\\
};
\addplot[ybar, draw=black, area legend, fill=exp] table[row sep=crcr] {%
3	0.0292211362975479\\
};
\addplot[ybar, draw=black, area legend, fill=estrada] table[row sep=crcr] {%
4	0.0414726700552392\\
};
\addplot[ybar, draw=black, area legend, fill=nbtexp] table[row sep=crcr] {%
5	0.0504451038575669\\
};
\end{axis}

\begin{axis}[%
width=0.328\columnwidth,
height=0.041\columnwidth,
at={(0.432\columnwidth,0.268\columnwidth)},
scale only axis,
xmin=-0.2,
xmax=14.2,
xtick={ 1,  2,  3,  4,  5,  6,  7,  8,  9, 10, 11, 12, 13},
ymin=0,
ymax=0.5,
axis background/.style={fill=white},
ylabel={$L$},
y label style={at={(1.05,.23)},rotate=-90,anchor=south,font=\footnotesize},
ticklabel style = {font=\tiny},
scaled y ticks=false,
yticklabel style={
        /pgf/number format/fixed,
        /pgf/number format/precision=2
},
]
\addplot[ybar, bar width=0.8, draw=black, area legend] table[row sep=crcr] {%
1	0.0416666666666667\\
2	0.0833333333333332\\
3	0.416666666666667\\
4	0.0833333333333333\\
};
\addplot[ybar, bar width=0.8, draw=black, area legend, fill=laplacian] table[row sep=crcr] {%
5	0.0416666666666667\\
};
\addplot[ybar, bar width=0.8, draw=black, area legend] table[row sep=crcr] {%
6	0.0416666666666667\\
7	0.0416666666666667\\
8	0.0416666666666667\\
9	0.0416666666666667\\
10	0.0416666666666667\\
11	0.0416666666666667\\
12	0.0416666666666667\\
13	0.0416666666666667\\
};
\end{axis}

\begin{axis}[%
width=0.328\columnwidth,
height=0.041\columnwidth,
at={(0.432\columnwidth,0.201\columnwidth)},
scale only axis,
xmin=-0.2,
xmax=14.2,
xtick={ 1,  2,  3,  4,  5,  6,  7,  8,  9, 10, 11, 12, 13},
ymin=0,
ymax=0.15907384894395,
axis background/.style={fill=white},
ticklabel style = {font=\tiny},
scaled y ticks=false,
yticklabel style={
        /pgf/number format/fixed,
        /pgf/number format/precision=2
},
ylabel={$\mathbb{L}(\operatorname{res})$},
y label style={at={(1.11,.23)},rotate=-90,anchor=south,font=\footnotesize},
]
\addplot[ybar, bar width=0.8, draw=black, area legend] table[row sep=crcr] {%
1	0.0582171338987795\\
2	0.0797345197910677\\
3	0.15907384894395\\
4	0.0797345197910676\\
};
\addplot[ybar, bar width=0.8, draw=black, area legend, fill=katz] table[row sep=crcr] {%
5	0.0582171338987794\\
};
\addplot[ybar, bar width=0.8, draw=black, area legend] table[row sep=crcr] {%
6	0.0706278554595445\\
7	0.0706278554595445\\
8	0.0706278554595445\\
9	0.0706278554595445\\
10	0.0706278554595445\\
11	0.0706278554595445\\
12	0.0706278554595445\\
13	0.0706278554595445\\
};
\end{axis}

\begin{axis}[%
width=0.328\columnwidth,
height=0.041\columnwidth,
at={(0.432\columnwidth,0.134\columnwidth)},
scale only axis,
xmin=-0.2,
xmax=14.2,
xtick={ 1,  2,  3,  4,  5,  6,  7,  8,  9, 10, 11, 12, 13},
ymin=0,
ymax=0.223456774752025,
axis background/.style={fill=white},
ticklabel style = {font=\tiny},
scaled y ticks=false,
yticklabel style={
        /pgf/number format/fixed,
        /pgf/number format/precision=2
},
ylabel={$\mathbb{L}(\exp)$},
y label style={at={(1.12,.23)},rotate=-90,anchor=south,font=\footnotesize},
]
\addplot[ybar, bar width=0.8, draw=black, area legend] table[row sep=crcr] {%
1	0.0292211362975481\\
2	0.0807770051554252\\
3	0.223456774752025\\
4	0.0807770051554251\\
};
\addplot[ybar, bar width=0.8, draw=black, area legend, fill=exp] table[row sep=crcr] {%
5	0.0292211362975479\\
};
\addplot[ybar, bar width=0.8, draw=black, area legend] table[row sep=crcr] {%
6	0.0695683677927535\\
7	0.0695683677927535\\
8	0.0695683677927535\\
9	0.0695683677927535\\
10	0.0695683677927535\\
11	0.0695683677927535\\
12	0.0695683677927535\\
13	0.0695683677927535\\
};
\end{axis}

\begin{axis}[%
width=0.328\columnwidth,
height=0.041\columnwidth,
at={(0.432\columnwidth,0.067\columnwidth)},
scale only axis,
xmin=-0.2,
xmax=14.2,
xtick={ 1,  2,  3,  4,  5,  6,  7,  8,  9, 10, 11, 12, 13},
ymin=0,
ymax=0.2,
axis background/.style={fill=white},
ticklabel style = {font=\tiny},
scaled y ticks=false,
yticklabel style={
        /pgf/number format/fixed,
        /pgf/number format/precision=2
},
ylabel={$\mathcal{L}$,\\$t_k = \exp(-k)$},
y label style={at={(1.5,.45)},rotate=-90,anchor=east,font=\footnotesize,align=left},
]
\addplot[ybar, bar width=0.8, draw=black, area legend] table[row sep=crcr] {%
1	0.0414726700552392\\
2	0.085697122563067\\
3	0.168927904866141\\
4	0.0856971225630669\\
};
\addplot[ybar, bar width=0.8, draw=black, area legend, fill=estrada] table[row sep=crcr] {%
5	0.0414726700552392\\
};
\addplot[ybar, bar width=0.8, draw=black, area legend] table[row sep=crcr] {%
6	0.0720915637371559\\
7	0.0720915637371559\\
8	0.0720915637371559\\
9	0.0720915637371559\\
10	0.0720915637371559\\
11	0.0720915637371559\\
12	0.0720915637371559\\
13	0.0720915637371559\\
};
\end{axis}

\begin{axis}[%
width=0.328\columnwidth,
height=0.041\columnwidth,
at={(0.432\columnwidth,0\columnwidth)},
scale only axis,
xmin=-0.2,
xmax=14.2,
xtick={ 1,  2,  3,  4,  5,  6,  7,  8,  9, 10, 11, 12, 13},
ymin=0,
ymax=0.1,
axis background/.style={fill=white},
ticklabel style = {font=\tiny},
scaled y ticks=false,
yticklabel style={
        /pgf/number format/fixed,
        /pgf/number format/precision=2
},
ylabel={$\mathbb{L}_1(\exp)$},
y label style={at={(1.3,.45)},rotate=-90,anchor=east,font=\footnotesize,align=left},
]
\addplot[ybar, bar width=0.8, draw=black, area legend] table[row sep=crcr] {%
1	0.0504451038575669\\
2	0.0845697329376854\\
3	0.0178041543026707\\
4	0.0845697329376854\\
};
\addplot[ybar, bar width=0.8, draw=black, area legend, fill=nbtexp] table[row sep=crcr] {%
5	0.0504451038575669\\
};
\addplot[ybar, bar width=0.8, draw=black, area legend] table[row sep=crcr] {%
6	0.0890207715133531\\
7	0.0890207715133531\\
8	0.0890207715133531\\
9	0.0890207715133531\\
10	0.0890207715133531\\
11	0.0890207715133531\\
12	0.0890207715133531\\
13	0.0890207715133531\\
};
\end{axis}
\end{tikzpicture}%
    \caption{Stationary distribution for the Markov chain induced by the different Laplacian. On the left panel, we have the probability mass at the target node at the end of the path, on the right one we report the different distributions. $\mathbb{L}(\operatorname{res})$ is used to denote $\mathbb{L}((1-\alpha x)^{-1})$ for $\alpha = \nicefrac{1}{2\rho(A)}$.}
    \label{fig:graph-with-trap}
\end{figure}

To describe the effect of the different formulations of the Laplacian $\mathbb{L}_\mu(f)$ for $\mu\in[0,1]$ and $f\in\mathcal{P}$, we consider the toy network $G_{5,8}$. Networks $G_{l,m}$, for $\ell,m=1,2,\ldots$ consist of a path with $l=2k+1$ vertices and of $m$ additional leaves all connected to the central node in the path. With this notation, the toy network considered is: 
\[
G_{5,8} =  \begin{tikzpicture}[baseline=(current bounding box.center),scale=1.5]
\Vertex[x=1.00,y=0.00,size=0.45,label={\tiny 1},color=white]{1}
\Vertex[x=2.00,y=0.00,size=0.45,label={\tiny 2},color=white]{2}
\Vertex[x=3.00,y=0.00,size=0.45,label={\tiny 3},color=white]{3}
\Vertex[x=4.00,y=0.00,size=0.45,label={\tiny 4},color=white]{4}
\Vertex[x=5.00,y=0.00,size=0.45,label={\tiny 5},color=white]{5}
\Vertex[x=3.38,y=0.32,size=0.45,label={\tiny 6},color=white]{6}
\Vertex[x=3.09,y=0.49,size=0.45,label={\tiny 7},color=white]{7}
\Vertex[x=2.75,y=0.43,size=0.45,label={\tiny 8},color=white]{8}
\Vertex[x=2.53,y=0.17,size=0.45,label={\tiny 9},color=white]{9}
\Vertex[x=2.53,y=-0.17,size=0.45,label={\tiny 10},color=white]{10}
\Vertex[x=2.75,y=-0.43,size=0.45,label={\tiny 11},color=white]{11}
\Vertex[x=3.09,y=-0.49,size=0.45,label={\tiny 12},color=white]{12}
\Vertex[x=3.38,y=-0.32,size=0.45,label={\tiny 13},color=white]{13}
\Edge(1)(2)
\Edge(2)(3)
\Edge(3)(4)
\Edge(3)(6)
\Edge(3)(7)
\Edge(3)(8)
\Edge(3)(9)
\Edge(3)(10)
\Edge(3)(11)
\Edge(3)(12)
\Edge(3)(13)
\Edge(4)(5)
\end{tikzpicture}.
\]
The cluster of nodes in the centre acts as a ``trap'' for a walker. An interesting dynamic would be one for which the steady state probability of having arrived at the end of the path is significant compared with that of having arrived at any other node. We report our results in Fig.~\ref{fig:graph-with-trap}.
We observe how the standard Laplacian $L$ concentrates most of the mass of its steady state at node $3$, the center of the trap. The same behavior is observed for all the other variants, except for the NBT $k$-path transformed Laplacian $\mathbb{L}_1(\operatorname{exp})$. This fits well with the model interpretation. Indeed, if a walker moves from node $3$, it is not possible to revisit it as backtracking is eliminated.

As a second example, consider an unlabeled tree generated uniformly at random using Wilf’s algorithm~\cite{MR640062}. To build it we use the implementation of the algorithm available in the Python \texttt{NetworkX} (v.3.3) library~\cite{paper:hagberg:2008} via a MATLAB interface. We construct a tree with $n = 100$ nodes and consider the probability matrices associated with the standard Laplacian $L$, the exponential $k$-walk transformed Laplacian $\mathbb{L}(\exp)$, the $k$-path transformed Laplacian $\mathcal{L}$ with $t_k = \exp(-\beta k)$ in~\eqref{eq:kpathlaplacian}, and exponential NBT $k$-walk transformed Laplacian $\mathbb{L}_{1}(\exp)$. With the associated probability matrices we simulate the connected Markov chain starting from the root node; see Section~\ref{sec:markov}. In Fig.~\ref{fig:exploring_tree} we report the evolution of the chain after $20$, $40$, and $80$ steps showing which nodes of the tree are explored by the different strategies. In all cases other than the classical Laplacian $L$, the results demonstrate the non-locality of the discrete diffusion process. 
\begin{figure}[b]
    \centering
    \stepcounter{figure}
    \subfloat[Exploration history of the different Markov chains on a tree graph with $n=100$ nodes.{We used orange to color the starting node and yellow to color the nodes visited in the exploration.}\label{fig:exploring_tree}]{\includegraphics[width=\columnwidth]{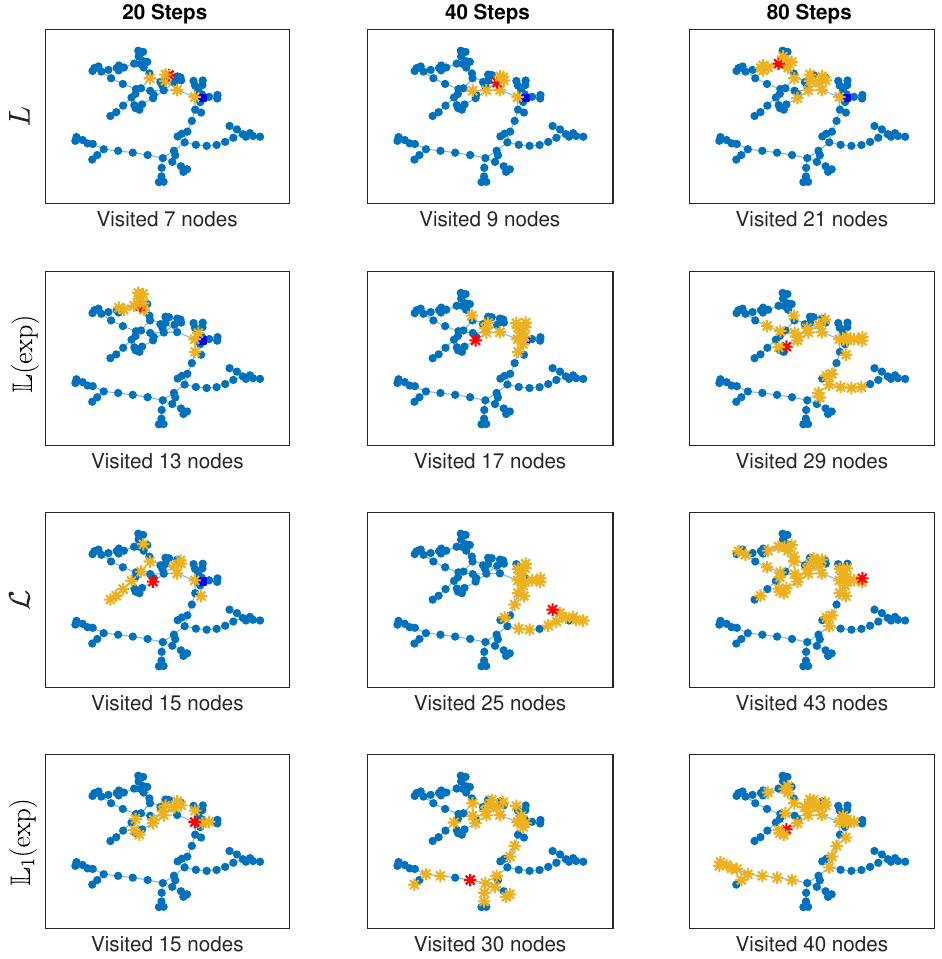}}
\end{figure}
\begin{figure}[t]
    \setcounter{subfigure}{1}
    \subfloat[Depiction of the spectral gap and eigenvalues for the probability matrices induced by the $k$-path exponentially transformed Laplacian $\mathcal{L}$, the exponential $k$-walk transformed Laplacian $\mathbb{L}(\exp)$ and its NBT counterpart $\mathbb{L}_1(\exp)$.\label{fig:tree-gap}]{\input{spectral_gap}}

    \addtocounter{figure}{-1} %
    \caption{Example of discrete-time diffusion on an unlabeled tree graph for Markov chains obtained through different formulations of the Laplacian.}
    \label{fig:tree-example}
\end{figure}
The $k$-path transformed Laplacian ($\mathcal{L}$) with exponential weights ($t_k=\exp(-k)$) and the exponential NBT $k$-walk transformed based Laplacian ($\mathbb{L}_1(\exp)$) are the ones that explore more nodes in the tree. To investigate the asymptotic behavior of the different chains we can also look at the gap between the eigenvalue of maximum modulus ($1$) and the one immediately below it, i.e., the \textit{spectral gap}. From Fig.~\ref{fig:tree-gap} we observe that the two Laplacians that explore the tree more widely have a larger gap compared to others.

\subsection{Average return probabilities}\label{sec:averagereturn}

In this section, we perform two types of experiments. For the smaller networks in Table~\ref{tab:complexnetworks}, we compute the average return probabilities (exactly) in order to investigate the behavior of different diffusion models. We then evaluate the performance of the randomized procedure in Alg.~\ref{alg:trace_estimation}.

\begin{figure}[htb]
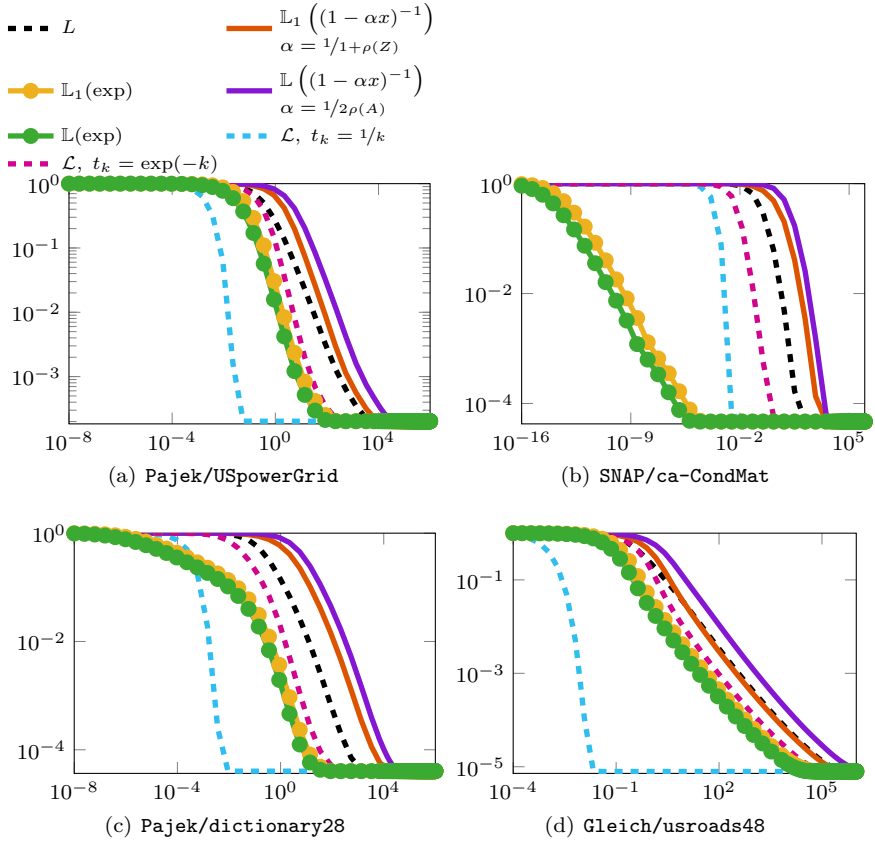

\centering

\subfloat[\texttt{Pajek/USpowerGrid}]{\input{PajekUSpowerGrid}}
\subfloat[\texttt{SNAP/ca-CondMat}\label{fig:avg-the-strange-case}]{\input{SNAPcaCondMat}}

\subfloat[\texttt{Pajek/dictionary28}]{\input{Pajekdictionary28}}
\subfloat[\texttt{Gleich/usroads48}\label{fig:usroads}]{\input{Gleichusroads48}}

\caption{Average return probabilities~\eqref{eq:average_return_probabilities} computed for matrices from Table~\ref{tab:complexnetworks} for which the computation of the path-based Laplacian was feasible for the dimension. Each subfigure displays results across six different Laplacian formulations on logarithmic time scales.}
\label{fig:average_return_probabilities}
\end{figure}

In Fig.~\ref{fig:average_return_probabilities} the graphs are separated based on the network considered. We plot 
the average return probability~\eqref{eq:average_return_probabilities} versus time on a logarithmic scale, 
comparing six different Laplacian formulations: the classical Laplacian $L$, the NBT variant based on Katz 
centrality $\mathbb{L}_1\left((1-\alpha x)^{-1}\right)$, the exponential NBT $k$-walk transformed Laplacian 
$\mathbb{L}_1(\exp)$, the Katz centrality $k$-walk Laplacian $\mathbb{L}\left((1-\alpha x)^{-1}\right)$, 
the exponential $k$-walk transformed Laplacian $\mathbb{L}(\exp)$, and two $k$-path transformed Laplacians 
$\mathcal{L}$ with different weight functions: $t_k = 1/k$ and $t_k = \exp(-k)$. The average return 
probabilities are computed exactly, i.e., through the eigenvalue decomposition of the relevant Laplacian matrix. 
The behavior observed is consistent across all considered networks. Notably, for exponential $k$-walk 
transformed Laplacians, the NBT variant $\mathbb{L}_1(\exp)$ exhibits a return probability that decays more 
slowly than $\mathbb{L}(\exp)$, confirming that removing backtracking walks decelerates the exploration process. 
In all cases except the one shown in Fig.~\ref{fig:avg-the-strange-case} for the network \texttt{SNAP/ca-CondMat}, 
the fastest convergence is achieved by the $\mathcal{L}$ transformed with weights $t_k = 1/k$, corresponding 
to $\beta = 1$ in~\eqref{eq:kpathlaplacian-global}. In all cases, $\mathcal{L}$ transformed with 
$t_k = \exp(-k)$ has slower convergence than $\mathbb{L}_\mu(\exp)$, with $\mu = 0,1$.

Another noteworthy finding is the behavior of the the $k$-walk transformed Laplacians $\mathbb{L}_\mu((1-\alpha x)^{-1})$ for $\mu = 0,1$. Both these result in a slower transition to the steady state compared to the classical Laplacian $L$. %

\begin{figure}[htb]
    \centering
    \input{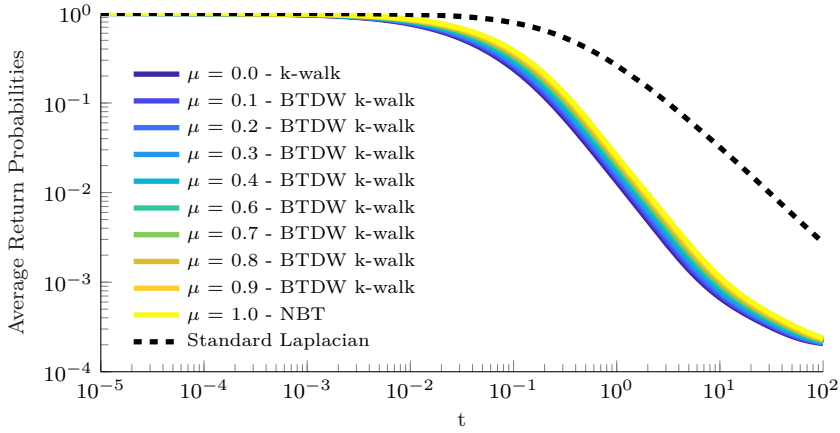}
    
    \caption{Average return probabilities~\eqref{eq:average_return_probabilities} computed for the \texttt{Gleich/usroads48} and the $\mathbb{L}_\mu(\exp)$ Laplacian for $0 \leq \mu \leq 1$ showing the effect of down-weighting more and more the NBT walks. The dashed line represents the average return probability of standard Laplacian $L$.}
    \label{fig:varying_mu}
\end{figure}

In Fig.~\ref{fig:varying_mu} we report the average return probability for the BTDW $k$-walk Laplacian $\mathbb{L}_\mu(\exp)$ as $\mu$ varies in $[0,1]$. 
The plot highlights the interpolation effect between the case $\mu = 0$, where the construction coincides with the $k$-walk transformed Laplacian $\mathbb{L}(\exp)$, and the case $\mu = 1$, which results in the NBT variant $\mathbb{L}_1(\exp)$.
We observe that suppressing backtracking walks slows down the overall exploration of the network, resulting in a higher average return probability compared to the case where all walks are counted, as in $\mathbb{L}(\exp)$. 
In other words, penalizing or removing immediate reversals limits the walker's ability to escape its local neighborhood, thereby reducing the effective diffusion rate across the graph. This highlights how the parameter $\mu$ controls the balance between local confinement and global exploration; smaller values promote faster spreading by allowing repeated edge traversals, while larger values enforce a more deliberate progression through the network.

\subsubsection{Randomized Average Return Probability Estimates}

In this section we test the idea of approximating the average return probability~\eqref{eq:average_return_probabilities} via the stochastic estimator \textsc{XNysTrace-exp}, as described in Section~\ref{sec:prob_estimation}. For the construction of the block rational Krylov space in Alg.~\ref{alg:trace_estimation} we use the \texttt{RKToolbox}~\cite{berljafa2014rational}.
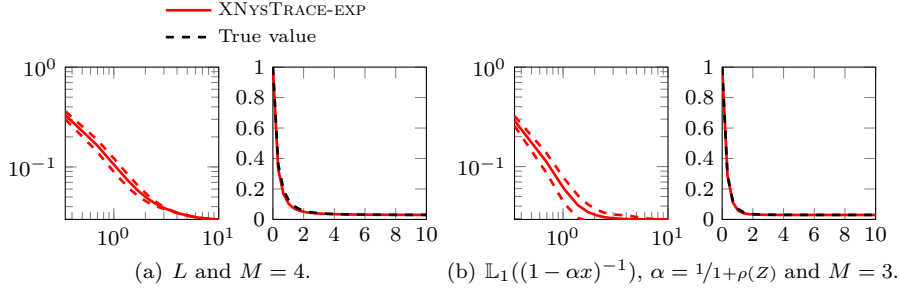
\begin{figure}[htbp]
    \centering
    \subfloat[$L$ and $M = 4$.\label{fig:trace_estimation-lap}]{\begin{tikzpicture}

\begin{axis}[%
width=0.17\columnwidth,
height=0.17\columnwidth,
at={(0\columnwidth,0\columnwidth)},
scale only axis,
xmin=0.344827586206897,
xmax=10,
ymin=0.0297853723516832,
ymax=1,
xmode=log,
ymode=log,
axis background/.style={fill=white},
legend style={legend cell align=left, align=left, draw=none,fill=none,font=\scriptsize, at={(1.1,1.3)}}
]
\addplot [color=red, line width=1.0pt]
  table[row sep=crcr]{%
0	1\\
0.344827586206897	0.3308976311394\\
0.689655172413793	0.166022765272814\\
1.03448275862069	0.101405577146512\\
1.37931034482759	0.072080820905094\\
1.72413793103448	0.0571507727079773\\
2.06896551724138	0.0487936670882285\\
2.41379310344828	0.0437465825361192\\
2.75862068965517	0.0404892048564293\\
3.10344827586207	0.0382491369256637\\
3.44827586206897	0.0366155724095531\\
3.79310344827586	0.0353655876242169\\
4.13793103448276	0.034375437529358\\
4.48275862068965	0.0335732806125813\\
4.82758620689655	0.0329144306581602\\
5.17241379310345	0.0323687997314118\\
5.51724137931035	0.031914631186757\\
5.86206896551724	0.0315353357470734\\
6.20689655172414	0.031217819998558\\
6.55172413793103	0.030951532772451\\
6.89655172413793	0.0307278674900641\\
7.24137931034483	0.0305397525385895\\
7.58620689655172	0.0303813504005219\\
7.93103448275862	0.0302478263256573\\
8.27586206896552	0.0301351655672431\\
8.62068965517241	0.0300400267546254\\
8.96551724137931	0.0299596232358572\\
9.31034482758621	0.0298916265545871\\
9.6551724137931	0.0298340876517526\\
10	0.0297853723516832\\
};

\addplot [color=red, dashed, line width=1.0pt, forget plot]
  table[row sep=crcr]{%
0	1\\
0.344827586206897	0.361330616988079\\
0.689655172413793	0.188742227572261\\
1.03448275862069	0.117220379644322\\
1.37931034482759	0.0825610110682844\\
1.72413793103448	0.0638487585607021\\
2.06896551724138	0.0529307226521608\\
2.41379310344828	0.0462129097941143\\
2.75862068965517	0.0419095195314626\\
3.10344827586207	0.039043027803027\\
3.44827586206897	0.0370494362616259\\
3.79310344827586	0.0355991805778287\\
4.13793103448276	0.0345001370126532\\
4.48275862068965	0.0336396022517707\\
4.82758620689655	0.0329496916357782\\
5.17241379310345	0.0323875831965273\\
5.51724137931035	0.0319246720463355\\
5.86206896551724	0.0315407276267317\\
6.20689655172414	0.0312207307671494\\
6.55172413793103	0.0309531133227262\\
6.89655172413793	0.0307287310957576\\
7.24137931034483	0.0305402274972652\\
7.58620689655172	0.0303816133763967\\
7.93103448275862	0.0302479729292833\\
8.27586206896552	0.0301352478585206\\
8.62068965517241	0.0300400732614873\\
8.96551724137931	0.0299596496946716\\
9.31034482758621	0.0298916417045632\\
9.6551724137931	0.0298340963794851\\
10	0.0297853774082686\\
};
\addplot [color=red, dashed, line width=1.0pt, forget plot]
  table[row sep=crcr]{%
0	1\\
0.344827586206897	0.300464645290721\\
0.689655172413793	0.143303302973367\\
1.03448275862069	0.0855907746487017\\
1.37931034482759	0.0616006307419037\\
1.72413793103448	0.0504527868552526\\
2.06896551724138	0.0446566115242962\\
2.41379310344828	0.0412802552781242\\
2.75862068965517	0.0390688901813959\\
3.10344827586207	0.0374552460483004\\
3.44827586206897	0.0361817085574802\\
3.79310344827586	0.035131994670605\\
4.13793103448276	0.0342507380460629\\
4.48275862068965	0.0335069589733919\\
4.82758620689655	0.0328791696805422\\
5.17241379310345	0.0323500162662964\\
5.51724137931035	0.0319045903271785\\
5.86206896551724	0.031529943867415\\
6.20689655172414	0.0312149092299666\\
6.55172413793103	0.0309499522221758\\
6.89655172413793	0.0307270038843706\\
7.24137931034483	0.0305392775799139\\
7.58620689655172	0.030381087424647\\
7.93103448275862	0.0302476797220313\\
8.27586206896552	0.0301350832759656\\
8.62068965517241	0.0300399802477634\\
8.96551724137931	0.0299595967770429\\
9.31034482758621	0.0298916114046111\\
9.6551724137931	0.0298340789240201\\
10	0.0297853672950978\\
};
\end{axis}

\begin{axis}[%
width=0.17\columnwidth,
height=0.17\columnwidth,
at={(0.23\columnwidth,0\columnwidth)},
scale only axis,
xmin=0,
xmax=10,
ymin=0,
ymax=1,
yminorticks=true,
axis background/.style={fill=white},
legend style={legend cell align=left, align=left, draw=none,fill=none,font=\scriptsize, at={(0.7,1.5)}}
]
\addplot [color=red, line width=1.0pt]
  table[row sep=crcr]{%
0	1\\
0.344827586206897	0.3308976311394\\
0.689655172413793	0.166022765272814\\
1.03448275862069	0.101405577146512\\
1.37931034482759	0.072080820905094\\
1.72413793103448	0.0571507727079773\\
2.06896551724138	0.0487936670882285\\
2.41379310344828	0.0437465825361192\\
2.75862068965517	0.0404892048564293\\
3.10344827586207	0.0382491369256637\\
3.44827586206897	0.0366155724095531\\
3.79310344827586	0.0353655876242169\\
4.13793103448276	0.034375437529358\\
4.48275862068965	0.0335732806125813\\
4.82758620689655	0.0329144306581602\\
5.17241379310345	0.0323687997314118\\
5.51724137931035	0.031914631186757\\
5.86206896551724	0.0315353357470734\\
6.20689655172414	0.031217819998558\\
6.55172413793103	0.030951532772451\\
6.89655172413793	0.0307278674900641\\
7.24137931034483	0.0305397525385895\\
7.58620689655172	0.0303813504005219\\
7.93103448275862	0.0302478263256573\\
8.27586206896552	0.0301351655672431\\
8.62068965517241	0.0300400267546254\\
8.96551724137931	0.0299596232358572\\
9.31034482758621	0.0298916265545871\\
9.6551724137931	0.0298340876517526\\
10	0.0297853723516832\\
};
\addlegendentry{\textsc{XNysTrace-exp}}

\addplot [color=black, line width=1.0pt, dashed]
  table[row sep=crcr]{%
0	1\\
0.344827586206897	0.362562058579189\\
0.689655172413793	0.195174602275109\\
1.03448275862069	0.123554616832822\\
1.37931034482759	0.0875749067807149\\
1.72413793103448	0.0676945514585937\\
2.06896551724138	0.0559198395413089\\
2.41379310344828	0.0485382464086643\\
2.75862068965517	0.0436786650516594\\
3.10344827586207	0.0403401186933859\\
3.44827586206897	0.0379605624930272\\
3.79310344827586	0.0362106955953636\\
4.13793103448276	0.0348898611891347\\
4.48275862068965	0.0338711895033474\\
4.82758620689655	0.0330716228443575\\
5.17241379310345	0.0324349927314917\\
5.51724137931035	0.0319221671029179\\
5.86206896551724	0.0315051407330916\\
6.20689655172414	0.0311633867347697\\
6.55172413793103	0.0308815400500801\\
6.89655172413793	0.030647886654858\\
7.24137931034483	0.0304533529098695\\
7.58620689655172	0.0302908132476021\\
7.93103448275862	0.0301546053771164\\
8.27586206896552	0.0300401838544473\\
8.62068965517241	0.0299438678788434\\
8.96551724137931	0.0298626545295109\\
9.31034482758621	0.0297940782887192\\
9.6551724137931	0.0297361038641109\\
10	0.0296870433491538\\
};
\addlegendentry{True value}

\end{axis}
\end{tikzpicture}%
}\hfil
    \subfloat[$\mathbb{L}_{1}((1-\alpha x)^{-1})$, $\alpha = \nicefrac{1}{1+\rho(Z)}$ and $M = 3$.
    \label{fig:trace_estimation-katzlap}]{\begin{tikzpicture}

\begin{axis}[%
width=0.17\columnwidth,
height=0.17\columnwidth,
at={(0\columnwidth,0\columnwidth)},
scale only axis,
xmin=0.344827586206897,
xmax=10,
ymin=0.0294121983144847,
ymax=1,
xmode=log,
ymode=log,
axis background/.style={fill=white},
legend style={legend cell align=left, align=left, draw=white!15!black}
]
\addplot [color=red, line width=1.0pt]
  table[row sep=crcr]{%
0	1\\
0.344827586206897	0.289364531753732\\
0.689655172413793	0.112621650828093\\
1.03448275862069	0.0586784620063966\\
1.37931034482759	0.0410838304710982\\
1.72413793103448	0.0349286838517004\\
2.06896551724138	0.0323882948689306\\
2.41379310344828	0.0311544857780596\\
2.75862068965517	0.0305130974767554\\
3.10344827586207	0.0301590943590421\\
3.44827586206897	0.0299395820715999\\
3.79310344827586	0.0297878552963677\\
4.13793103448276	0.029678072276983\\
4.48275862068965	0.0295985048958929\\
4.82758620689655	0.0295415808586745\\
5.17241379310345	0.029501418988615\\
5.51724137931035	0.0294733948737358\\
5.86206896551724	0.0294539952930643\\
6.20689655172414	0.0294406398627094\\
6.55172413793103	0.0294314799300319\\
6.89655172413793	0.0294252134367695\\
7.24137931034483	0.0294209337116635\\
7.58620689655172	0.0294180141941892\\
7.93103448275862	0.0294160240913539\\
8.27586206896552	0.0294146682158211\\
8.62068965517241	0.029413744759856\\
8.96551724137931	0.0294131159664352\\
9.31034482758621	0.0294126878980122\\
9.6551724137931	0.0294123965446388\\
10	0.0294121983144847\\
};

\addplot [color=red, dashed, line width=1.0pt, forget plot]
  table[row sep=crcr]{%
0	1\\
0.344827586206897	0.322427919294498\\
0.689655172413793	0.137861696795772\\
1.03448275862069	0.0751210875772238\\
1.37931034482759	0.051532240687751\\
1.72413793103448	0.0417131723892041\\
2.06896551724138	0.0370562211066318\\
2.41379310344828	0.0346821822006035\\
2.75862068965517	0.0334998805768338\\
3.10344827586207	0.0329322994518346\\
3.44827586206897	0.0326187209545166\\
3.79310344827586	0.0323561692942576\\
4.13793103448276	0.0320591823154915\\
4.48275862068965	0.0317164648541582\\
4.82758620689655	0.0313518499572515\\
5.17241379310345	0.0309962521332016\\
5.51724137931035	0.0306733468961899\\
5.86206896551724	0.0303958036715895\\
6.20689655172414	0.0301669429113363\\
6.55172413793103	0.0299840098044233\\
6.89655172413793	0.0298411578250965\\
7.24137931034483	0.029731536585736\\
7.58620689655172	0.0296485108912759\\
7.93103448275862	0.0295862454808533\\
8.27586206896552	0.0295398966156819\\
8.62068965517241	0.0295055915945949\\
8.96551724137931	0.0294803121109781\\
9.31034482758621	0.0294617474975175\\
9.6551724137931	0.0294481513827547\\
10	0.0294382161896584\\
};
\addplot [color=red, dashed, line width=1.0pt, forget plot]
  table[row sep=crcr]{%
0	1\\
0.344827586206897	0.256301144212966\\
0.689655172413793	0.0873816048604138\\
1.03448275862069	0.0422358364355695\\
1.37931034482759	0.0306354202544454\\
1.72413793103448	0.0281441953141966\\
2.06896551724138	0.0277203686312294\\
2.41379310344828	0.0276267893555157\\
2.75862068965517	0.0275263143766769\\
3.10344827586207	0.0273858892662496\\
3.44827586206897	0.0272604431886831\\
3.79310344827586	0.0272195412984779\\
4.13793103448276	0.0272969622384745\\
4.48275862068965	0.0274805449376276\\
4.82758620689655	0.0277313117600976\\
5.17241379310345	0.0280065858440285\\
5.51724137931035	0.0282734428512816\\
5.86206896551724	0.0285121869145392\\
6.20689655172414	0.0287143368140825\\
6.55172413793103	0.0288789500556405\\
6.89655172413793	0.0290092690484425\\
7.24137931034483	0.029110330837591\\
7.58620689655172	0.0291875174971024\\
7.93103448275862	0.0292458027018545\\
8.27586206896552	0.0292894398159604\\
8.62068965517241	0.0293218979251171\\
8.96551724137931	0.0293459198218923\\
9.31034482758621	0.0293636282985069\\
9.6551724137931	0.0293766417065228\\
10	0.0293861804393111\\
};
\end{axis}

\begin{axis}[%
width=0.17\columnwidth,
height=0.17\columnwidth,
at={(0.23\columnwidth,0\columnwidth)},
scale only axis,
xmin=0,
xmax=10,
ymin=0,
ymax=1,
yminorticks=true,
axis background/.style={fill=white},
legend style={legend cell align=left, align=left, draw=white!15!black}
]
\addplot [color=red, line width=1.0pt]
  table[row sep=crcr]{%
0	1\\
0.344827586206897	0.289364531753732\\
0.689655172413793	0.112621650828093\\
1.03448275862069	0.0586784620063966\\
1.37931034482759	0.0410838304710982\\
1.72413793103448	0.0349286838517004\\
2.06896551724138	0.0323882948689306\\
2.41379310344828	0.0311544857780596\\
2.75862068965517	0.0305130974767554\\
3.10344827586207	0.0301590943590421\\
3.44827586206897	0.0299395820715999\\
3.79310344827586	0.0297878552963677\\
4.13793103448276	0.029678072276983\\
4.48275862068965	0.0295985048958929\\
4.82758620689655	0.0295415808586745\\
5.17241379310345	0.029501418988615\\
5.51724137931035	0.0294733948737358\\
5.86206896551724	0.0294539952930643\\
6.20689655172414	0.0294406398627094\\
6.55172413793103	0.0294314799300319\\
6.89655172413793	0.0294252134367695\\
7.24137931034483	0.0294209337116635\\
7.58620689655172	0.0294180141941892\\
7.93103448275862	0.0294160240913539\\
8.27586206896552	0.0294146682158211\\
8.62068965517241	0.029413744759856\\
8.96551724137931	0.0294131159664352\\
9.31034482758621	0.0294126878980122\\
9.6551724137931	0.0294123965446388\\
10	0.0294121983144847\\
};

\addplot [color=black, line width=1.0pt, dashed]
  table[row sep=crcr]{%
0	1\\
0.344827586206897	0.28982713237951\\
0.689655172413793	0.123936717527597\\
1.03448275862069	0.0691999854431815\\
1.37931034482759	0.0480287039660786\\
1.72413793103448	0.0389240666064195\\
2.06896551724138	0.0346413516100531\\
2.41379310344828	0.032458677229138\\
2.75862068965517	0.0312662944945071\\
3.10344827586207	0.0305768634037338\\
3.44827586206897	0.0301603550719883\\
3.79310344827586	0.0299004021920711\\
4.13793103448276	0.0297342924193281\\
4.48275862068965	0.029626343915644\\
4.82758620689655	0.0295553414300731\\
5.17241379310345	0.0295082332150501\\
5.51724137931035	0.0294767807318588\\
5.86206896551724	0.0294556836728744\\
6.20689655172414	0.0294414839152237\\
6.55172413793103	0.0294319018241745\\
6.89655172413793	0.0294254231034804\\
7.24137931034483	0.0294210360979162\\
7.58620689655172	0.0294180620486006\\
7.93103448275862	0.0294160440681306\\
8.27586206896552	0.0294146738516904\\
8.62068965517241	0.0294137429596548\\
8.96551724137931	0.0294131102611831\\
9.31034482758621	0.0294126800890606\\
9.6551724137931	0.0294123875357479\\
10	0.0294121885321424\\
};

\end{axis}
\end{tikzpicture}%
}
    
    \caption{Randomized approximation of the average return probabilities~\eqref{eq:average_return_probabilities} with the \textsc{XNysTrace-exp} algorithm (Alg.~\ref{alg:trace_estimation}) on network \texttt{Newman/karate} and $M$ sample vectors. The dashed lines in the left panels of Fig.~\ref{fig:trace_estimation-lap} and~\ref{fig:trace_estimation-katzlap} depicts the error estimate obtained from Alg.~\ref{alg:trace_estimation}. In the right panels we overlap the estimate obtained from the \textsc{XNysTrace-exp} in Alg.~\ref{alg:trace_estimation} and the exact value computed from the eigenvalue decomposition.}
    \label{fig:trace_estimation}
\end{figure}
In Fig.~\ref{fig:trace_estimation-lap} and~\ref{fig:trace_estimation-katzlap} we report the results obtained for the estimation of the average probability of return using the algorithm on a small-sized network,  \texttt{Newman/karate} with $n = 32$ nodes, for which we can compute it exactly and compare the results. Due to the small dimension, we use few sampling vectors in the two cases, $M=4$ and $M=3$ respectively. As can be observed, the agreement between the stochastic estimate and the correct value is good enough and in any case adequate to describe the behavior of the two diffusion models.

\begin{figure}[htbp]
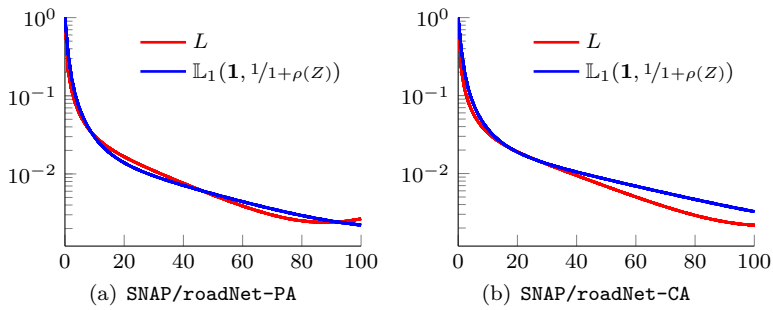

    \centering
    \subfloat[\texttt{SNAP/roadNet-PA}\label{fig:roadnet-pa}]{\input{roadnet-pa-random}}
    \subfloat[\texttt{SNAP/roadNet-CA}\label{fig:roadnet-ca}]{\input{roadnet-ca-random}}
    
    \caption{Randomized approximation of the average return probabilities~\eqref{eq:average_return_probabilities} with the \textsc{XNysTrace-exp} algorithm (Alg.~\ref{alg:trace_estimation}). Each experiment is run 10 times, with curves plotted using the same color and stroke thickness per repetition.}
    \label{fig:large-randomized}
\end{figure}

As a second example, consider two complex networks from Table~\ref{tab:complexnetworks}, for which exact computation of the average return probability is infeasible due to memory limitations (\texttt{SNAP/roadNet-PA} and \texttt{SNAP/roadNet-CA}, with $n=1087562$ and $n=1957027$ nodes, respectively). We select $M=30$ sampling vectors and apply Algorithm~\ref{alg:trace_estimation} over the interval $[0,t^*] = [0,100]$ with $n_t = 90$ equally spaced nodes. The estimation is repeated ten times, with each repetition plotted in Fig.~\ref{fig:large-randomized} using consistent stroke and color across curves.

As in the previous case, we focus on the standard Laplacian $L$ as a reference, along with the resolvent NBT $k$-walk transformed Laplacian $\mathbb{L}_1((1-\alpha x)^{-1})$. For this, we set $\alpha = \nicefrac{1}{1+\rho(Z)}$. The GMRES algorithm used in the rational Krylov subspace method is configured to achieve a tolerance of $10^{-6}$.

The results in Fig.s~\ref{fig:roadnet-pa} and~\ref{fig:roadnet-ca} for networs \texttt{SNAP/roadNet-PA} and \texttt{SNAP/roadNet-CA} can be compared to those in Fig.~\ref{fig:usroads} for the \texttt{Gleich/usroads48} network. Both networks represent road systems, suggesting a similar reciprocal behavior between the standard and NBT Laplacians, which is observed within the error margin due to randomization. For both networks, the \texttt{AAA} algorithm yields a rational approximation with 13 poles, and the internal GMRES method converges to the required tolerance with iteration counts between 6 and 12, depending on the pole.

\subsection{Matrix-vector products with resolvent $k$-walk transformed Laplacian}\label{sec:katzexperiment}

\begin{figure}[htbp]
    \centering
    \includegraphics[width=\columnwidth]{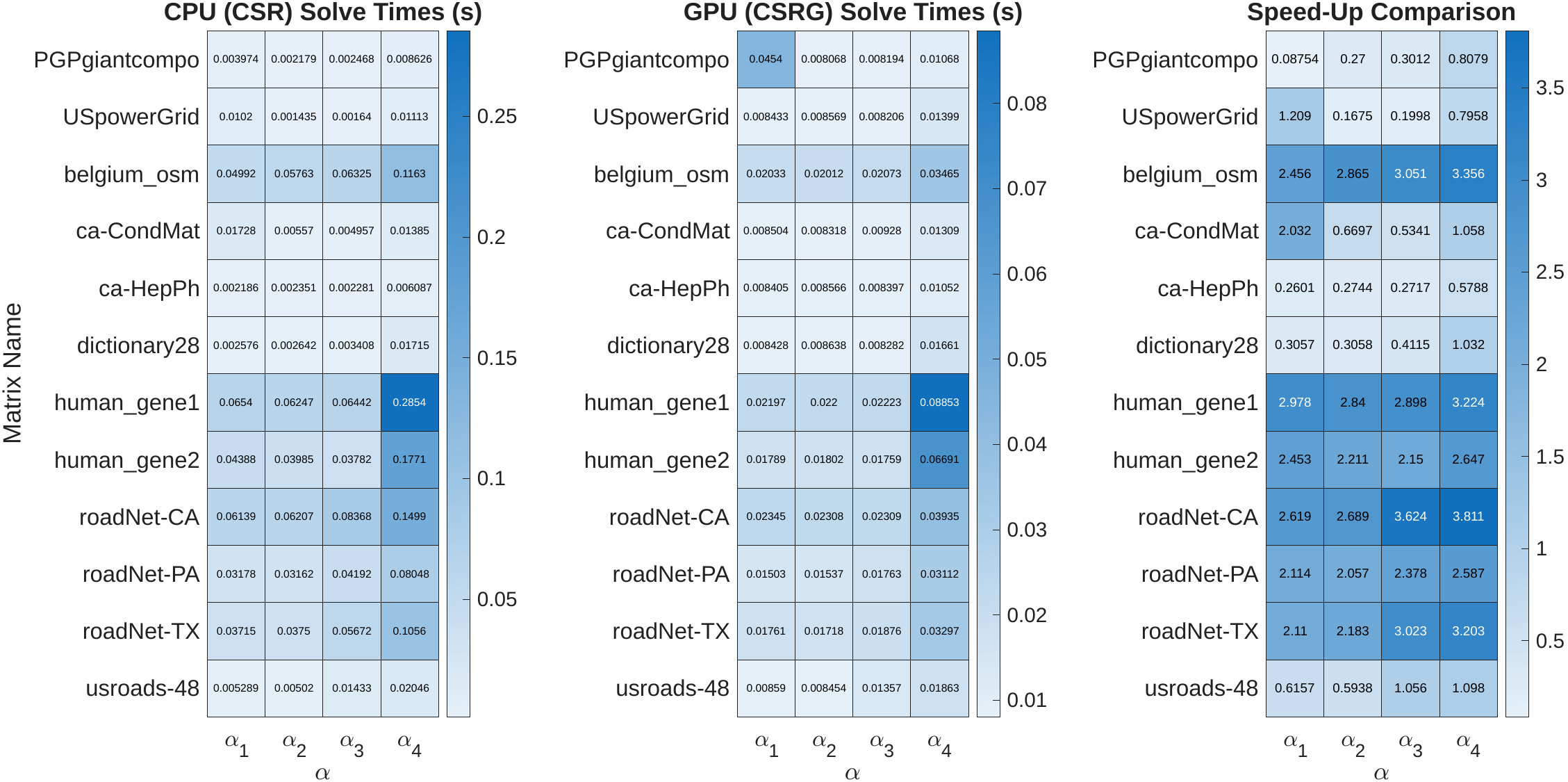}    
    \caption{The two heatmaps show the time required to apply a matrix-vector product with $\mathbb{L}_{\mu}((1 - \alpha x)^{-1})$. Four equally spaced values of $\alpha$ logarithmically scaled between $10^{-3}$ and $10^{-\nicefrac{1}{2}}$, based on the dominant eigenvalue of matrix $Z$ (see Proposition~\ref{pro:whe-we-converge}). Rows correspond to the twelve complex networks detailed in Table~\ref{tab:complexnetworks}.}
    \label{fig:distributed_and_gputed_katz}
\end{figure}

Here, we conduct tests on the solution of linear systems involving the matrix $\mathcal{A}(\alpha)$ across several examples of real-world networks. For each network, we restrict our analysis to its largest connected component. The parameter $\alpha$ is selected as four equally spaced, logarithmically scaled values between $10^{-3}$ and $10^{-\nicefrac{1}{2}}$, scaled according to the dominant eigenvalue of the matrix $Z$ in Proposition~\ref{pro:whe-we-converge}. The Conjugate Gradient algorithm, preconditioned with the AMG method from Section~\ref{sec:solving_katz}, is employed with a relative tolerance of $\varepsilon = 10^{-9}$. To minimize the construction time of the multigrid hierarchy, OpenMP parallelization is used with $64$ threads. {Results are displayed in Figure~\ref{fig:distributed_and_gputed_katz}, where we display the time required to apply a matrix-vector product with $\mathbb{L}_{\mu}((1 - \alpha x)^{-1})$, with $\alpha$ selected as described above.}

    For each of the twelve networks  in Table~\ref{tab:complexnetworks}, both CPU and GPU execution times vary only marginally as $\alpha$ changes.  This indicates that the matrix
    structure induced by the resolvent operator does not significantly affect algorithmic stability or solver convergence.  
    In practice, solving the
    resolvent-regularized system is nearly as efficient as performing a small number of standard sparse matrix-vector multiplications with the
    original graph Laplacian $L$.  This demonstrates the effectiveness of the AMG preconditioning strategy~\cite{DAMBRA2023100463} and the scalability of the approach across
    a wide range of graph sizes and topologies.
    The GPU implementation achieves speedups exceeding one order of magnitude for several cases, particularly on larger and denser
    networks such as \texttt{roadNet-CA}, \texttt{roadNet-PA}, \texttt{human\_gene2}, and \texttt{usroads-48}.  Even in instances where
    acceleration is less dramatic, GPU solve times remain comparable to or {are} faster than CPU computation.  This confirms that reformulating the
    sparse operator into the CSRG format allows efficient utilization of parallel devices.  Networks with higher edge counts benefit
    most, as the available parallelism enables large batches of row operations to proceed concurrently.

    Overall, the experiments demonstrate that applying the resolvent-based $k$-walk transformed Laplacian is computationally practical at scale.  The combination of AMG
preconditioning and CSRG-based GPU kernels yields substantial performance gains without sacrificing numerical robustness, making the method well
suited for large network analysis tasks where repeated linear solves are required.

\section{Conclusions and outlook}

In this paper, we introduced a novel framework for constructing diffusion operators on complex networks, extending the traditional Laplacian operator through the use of graph-based walks. This approach enables a flexible and nuanced interpretation of network centrality; while the classical Laplacian operator aligns with degree centrality, our generalized construction leverages a generic total communicabiltiy centrality, facilitating the use of diverse matrix function-based centrality measures to shape the diffusion dynamics. In our framework we have treated also the inclusion of backtrack-downweighted and nonbacktracking walks, which incorporate memory into the diffusion process thus enhancing its descriptive power.

We proposed and analyzed efficient computational strategies for these operators, along with methods to compute key quantities such as the average return probability of the diffusion process. Demonstrating applicability on real-world networks, we showed how our approach differs from existing path-based diffusion operators, highlighting the distinctive properties and flexibility of this new class of operators. Our framework offers a valuable tool for adapting diffusion dynamics across various network contexts.
Future work will focus on new directions in network analysis and modeling, such as the possibility to consider and generalize the discrete version of differential equations beyond diffusion (e.g., wave equation or Schr\"{o}dinger equation). From a computational viewpoint, we also plan to integrate the tools contained in the \texttt{PSCToolkit} library to be able to generate and treat these problems in a systematic way for large networks on distributed machines equipped with hardware accelerators such as GPUs.

\begin{acknowledgements}
The authors would like to thank the anonymous reviewer for their constructive feedback, which helped improve this paper.
\end{acknowledgements}

\small{\noindent\rmfamily\textbf{Data and reproducibility} All matrices used in this work are drawn from the \texttt{SuiteSparse Matrix Collection}~\cite{suitesparse} (\href{https://sparse.tamu.edu/}{sparse.tamu.edu}). The code required to reproduce all experiments and figures is publicly available at 
\href{https://github.com/Cirdans-Home/walk-laplacian}{github.com/Cirdans-Home/walk-laplacian}.}

\bibliographystyle{spmpsci}
\bibliography{biblionbt}%

\end{document}